\documentclass[journal, onecolumn, draftcls,12pt]{IEEEtran}
\usepackage{amssymb,amsmath,epsfig,graphicx,theorem}
\usepackage{amsfonts}
\usepackage{bm}
\usepackage{arydshln}
\usepackage[linesnumbered,ruled,vlined]{algorithm2e}
\newtheorem{Theo}{Theorem}
\newtheorem{Lem}{Lemma}

\newtheorem{Def}{Definition}
\newtheorem{Remark}{Remark}

\SetKwInput{KwInput}{Input}                
\SetKwInput{KwOutput}{Output}              
\SetKwInput{Initialization}{Initialization}  

\usepackage{epsfig,rotating,setspace,latexsym,amsmath,epsf,amssymb,bm,color}
\usepackage{cite}

\IEEEoverridecommandlockouts

\title{The Capacity of Private Information Retrieval Under Arbitrary Collusion Patterns}

\author{Xinyu Yao, Nan Liu, Wei~Kang%
\thanks{X. Yao and W. Kang are with the School of Information Science and Engineering,
Southeast University, Nanjing, China (email: \{xyyao,wkang\}@seu.edu.cn). N. Liu is with the National Mobile Communications Research Laboratory,
Southeast University, Nanjing, China (email: nanliu@seu.edu.cn).}%
}

\begin{document}

\maketitle
\begin{abstract}
We study the private information retrieval (PIR) problem  under arbitrary collusion pattern for replicated databases. We find its capacity, which is the same as the capacity of the original PIR problem with the number of databases $N$ replaced by a number $S^*$, which is the optimal solution to a linear programming problem that is a function of the collusion pattern. Hence, the collusion pattern affects the capacity of the PIR problem only through the number $S^*$. 
\end{abstract}

\section{introduction}
The problem of private information retrieval (PIR) was first proposed in \cite{chor1995private}, where the user wants to retrieve \emph{a certain bit} out of $K$ bits from $N$ replicated databases without revealing which bit is of interest to any single database. The design objective in \cite{chor1995private} is to minimize the upload cost and the download cost between the user and the databases. 
The PIR problem was reformulated in \cite{sun2017capacity} from an information-theoretic perspective, where  the user wants to retrieve a \emph{sufficiently large} message from the databases so that the download cost is minimized. This problem was fully solved by Sun and Jafar \cite{sun2017capacity},  where the capacity of the PIR problem was shown to be
\begin{align}
C_{\text{PIR}}=\left(1+\frac{1}{N}+\frac{1}{N^2}+\cdots+\frac{1}{N^{K-1}} \right)^{-1},   \label{CapacityPIR}
\end{align}
which is defined as the ratio of the size of the desired message to the total number of downloaded symbols from the databases.  
The capacity increases with the number of databases $N$, since with the help of more databases, the privacy of the user can be hidden better from any single database. Many interesting extensions and variations for the PIR problem have since then been studied \cite{sun2018colluding, sun2018private, zhang2017general, tajeddine2017arbitrary, zhang2017private, tajeddine2018robust, tajeddine2019byzantine, Holzbaur2019linear, 
wang2017linear, 
wang2018secure, wang2017secure, jia2017disjoint, 
wang2019symmetric, 
wang2019adversaries, zhang2019subpacket, Penas2019local, tajeddine2019colluding, 
banawan2018capacity2, sun2018capacity2, tajeddine2017robust, banawan2018capacity, tajeddine2018private, lin2018mds, tandon2017capacity, wei2018fundamental, wei2018cache, chen2017capacity, heidarzadeh2018capacity, abdul2017private, li2018single, banawan2018multi, shariatpanahi2018multi, banawan2018asymmetry, lin2018asymmetry, banawan2018noisy, banawan2018private, tian2018capacity, tajeddine2018private2, d2018lifting, sun2017optimal, kim2017cache, yang2018private, kumar2018private, raviv2018private, 
wang2018epsilon, 
Hsuan2019Weakly, heidarzadeh2019single, kazemi2019single, heidarzadeh2018capacityITW, banawan2019capacity, zhu2019new, zhou2019capacity, woolsey2019optimal, jia2019asymptotic, kazemi2019private, kadhe2019equivalence, sun2019breaking, jia2019x, sun2018multiround, OurArXiv, banawan2019improved, 
wei2019capacity, wei2018private, kadhe2017private, vajha2017binary, wang2018capacity, kumar2019achieving, jia2019cross, zhang2019private, banawan2018private2, shah2014one, fanti2014multi, melchor2008fast, chan2015private, lavauzelle2018private, blackburn2017pir, attia2018capacity, blackburn2019pir, fazeli2015codes, tian2018shannon, xu2018building, lin2019improved, wang2019mismatched, chee2019generalization, samy2019on, banawan2018private3, wei2018capacity, xu2019capacity, zhang2019access, Li2019repair, Guo2019leak, Xu2018Subpacket, Heidarzadeh201910, Tian201910, Lieb201911, Wang201912}.

One of the first variations studied was that of the colluding databases \cite{sun2018colluding}, where some subsets of databases may communicate and collude to learn about the message index that is of interest to the user. To preserve privacy under possible collusion among databases, the number of downloaded symbols needs to be increased. The first study on database collusion focused on the case where we have replicated databases, i.e., each database stores a replica of the entirety of the $K$ files, and $T$-colluding databases, where it is assumed that up to $T$ number of databases may collude. Sun and Jafar \cite{sun2018colluding} proved that the capacity of the $T$-colluding PIR problem for replicated databases, is
\begin{align}
C_{\text{PIR}}=\left(1+\frac{T}{N}+\left(\frac{T}{N} \right)^2+\cdots+\left(\frac{T}{N} \right)^{K-1} \right)^{-1}.   \label{CapacityTPIR}
\end{align}
Comparing (\ref{CapacityTPIR}) with (\ref{CapacityPIR}), we see that when any $T$ databases may collude, the number of effective databases has decreased from $N$ to $\frac{N}{T}$, where $\frac{N}{T}$ does not need to be an integer. 

Following \cite{sun2018colluding}, many extensions of $T$-colluding PIR have been studied \cite{wang2017secure, wang2017linear, tajeddine2017arbitrary, jia2017disjoint, zhang2017general, zhang2017private, wang2018secure, sun2018private, banawan2018capacity2, tajeddine2018robust, wang2019symmetric, tajeddine2019byzantine, wang2019adversaries, zhang2019subpacket, Penas2019local, Holzbaur2019linear, tajeddine2019colluding}, among which MDS-coded databases with $T$-colluding generated a lot of research interest \cite{tajeddine2017arbitrary, tajeddine2018robust, sun2018private, tajeddine2019byzantine, Holzbaur2019linear, zhang2017general, zhang2017private}. The MDS-coded databases scenario is the case where the messages are encoded using an $[N,J]$ MDS code, and the coded bits are stored in the $N$ databases. Unlike the replicated databases scenario, where each database has the ability to reconstruct all $K$ messages, here, any $J$ databases together can reconstruct the $K$ messages. Thus, the replicated databases scenario is a special case of the MDS-coded databases scenario when $J=1$. Finding the capacity of the $T$-colluding PIR problem with MDS-coded databases is difficult, and remains open in general \cite{sun2018private, Holzbaur2019linear}. 

While most works focused on the $T$-colluding structure of the databases, where any up to $T$ databases may collude, it is of interest to study more general collusion patterns due to the possible heterogeneity of the databases. An arbitrary collusion pattern may be represented by its maximal colluding sets \cite{tajeddine2017arbitrary, zhang2017private} as $\mathcal{P}=\{\mathcal{T}_1, \mathcal{T}_2, \cdots, \mathcal{T}_M\}$, where the databases in set $\mathcal{T}_m$, $m \in [1:M]$ may collude, and there are $M$ such colluding sets. Tajeddine \emph{et. al}  \cite{tajeddine2017arbitrary}  proposed  the PIR problem under arbitrary collusion patterns and studied it  for MDS-coded databases. Several other works followed, including \cite{jia2017disjoint} for replicated databases, \cite[Section VII]{zhang2017private} for MDS-coded databases, and some discussions in \cite[Appendix D]{sun2018private}, for both the replicated and MDS-coded databases scenarios.

In this paper, we focus on the PIR problem under arbitrary collusion patterns for the replicated databases scenario. The known results for this problem thus far is
1) the capacity for the special case of disjoint colluding sets\cite{jia2017disjoint};
2) the capacity for the special case of cyclically contiguous databases \cite[Appendix D]{sun2018private};
3) 
 a rate of (\ref{CapacityTPIR}) is achievable for $T \triangleq\max_{\mathcal{T} \in \mathcal{P}} |\mathcal{T}|$, i.e., we may consider the more strict collusion pattern where any up to the maximum number of colluding databases in $\mathcal{P}$ may collude. This is also the result we obtain when specializing \cite{zhang2017private} to the replicated databases scenario;
4) 
a rate indicated by Theorem 2 in \cite{tajeddine2017arbitrary},  specialized to the replicated databases scenario by setting $k=1$, is achievable.  
As can be seen, the understanding of the PIR problem under arbitrary collusion patterns for replicated databases is still rather limited. 

In this paper, we find the PIR capacity under arbitrary collusion patterns for the replicated databases scenario. Though collusion patterns are diverse, and at first glance, the problem requires a case-by-case analysis due to the property of each specific collusion pattern \cite{zhang2017private}, we provide a general formula for the PIR capacity that holds true for any collusion pattern $\mathcal{P}$. The capacity formula is shown to be
\begin{align}
C_\mathcal{P}=\left(1+\frac{1}{S^*}  +\left(\frac{1}{S^{^*}}\right)^2+\cdots+\left(\frac{1}{S^{*}} \right)^{K-1} \right)^{-1},  \label{CapacityArb01}
\end{align}
where $S^*$ is the optimal value of the following linear programming problem
 \begin{align}
\max_{\mathbf{y}} \quad & \mathbf{1}_{N}^T \mathbf{y} \nonumber\\
 \text{subject to} \quad & \mathbf{B}^T_\mathcal{P} \mathbf{y} \leq\mathbf{1}_M \nonumber\\
 & \mathbf{y} \geq \mathbf{0}_N, \nonumber
 \end{align}
 where $\mathbf{B}_\mathcal{P}$ is the incidence matrix, of size $N \times M$, of the collusion pattern $\mathcal{P}$, i.e., if DB $n$ is in the $m$-th colluding set $\mathcal{T}_m$ in $\mathcal{P}$, we let the $(n,m)$-th element of $\mathbf{B}_\mathcal{P}$ be $1$, otherwise, it is zero. $\mathbf{1}_k$ ($\mathbf{0}_k$) is the column vector of size $k$ whose elements are all one (zero). 
Comparing (\ref{CapacityArb01}) with (\ref{CapacityPIR}) and (\ref{CapacityTPIR}), we find that the number of effective databases under arbitrary collusion pattern $\mathcal{P}$ is $S^*$ which is related to the collusion pattern $\mathcal{P}$ through a linear programming solution. 

The difficulty of finding the capacity of the PIR problem under arbitrary collusion patterns for replicated databases comes from finding a \emph{common} proof and capacity expressions that works for \emph{any} collusion pattern. Towards this end, the tools and ideas that we use in proving the capacity result include 1) using the sub-modular property of the entropy function\cite{schrijver2003combinatorial} to prove a general inequality, which is used in place of Han's inequality for $T$-colluding \cite{sun2018colluding},  for the induction argument of the converse; 2) linking the achievable PIR rate and its converse to the optimal solution of two linear programming problems; 3) using the duality of linear programming problems to show that the achievability and converse results meet, yielding the capacity. 

\section{system model} \label{secSystemModel}
Consider the problem where $K$ messages are stored on $N$ replicated databases. The $K$ messages, denoted as $W_1, \cdots, W_K$, are independent and each message consists of $L$ symbols, which are independently and uniformly distributed over a finite field $\mathbb{F}_q$, where $q$ is the size of the field, i.e.,
\begin{align}
H(W_k)&=L, \qquad k=1,...,K, \label{LL}\\
H(W_1,...,W_K)&=H(W_1)+H(W_2)+\cdots +H(W_K). \nonumber
\end{align}

A user wants to retrieve message $W_\theta$, $\theta \in [1:K]$, by sending designed queries to the databases, where the query sent to the $n$-th databese is denoted as $Q_n^{[\theta]}$. Since the queries are designed by the user, who do not know the content of the messages, we have
\begin{align}
I(W_{1:K}; Q_{1:N}^{[\theta]})=0, \quad \forall \theta \in [1:K]. \label{SysInd}
\end{align}
Upon receiving the query $Q_{n}^{[\theta]}$, Database $n$ calculates the answer, denoted as $A_{n}^{[\theta]}$, based on the query received $Q_{n}^{[\theta]}$ and the messages $W_{1:K}$, i.e.,
\begin{align}
H(A_{n}^{[\theta]}|Q_{n}^{[\theta]},W_{1:K})=0, \quad \forall n \in [1:N], \theta \in [1:K]. \label{DatabaseHonest}
\end{align}

The queries need to be designed such that the user is able to reconstruct the desired message $W_\theta$ 
from all the answers received from the databases, i.e., 
\begin{align}
H(W_\theta|A_{1:N}^{[\theta]},Q_{1:N}^{[\theta]})=0,  \quad \forall \theta \in [1:K]. \label{NoError}
\end{align}
The queries also need to be designed such that the privacy of the user is preserved. In this paper, we consider colluding databases, and furthermore, the collusion pattern can be arbitrary. We represent the collusion pattern as $\mathcal{P}=\{\mathcal{T}_1, \mathcal{T}_2, \cdots, \mathcal{T}_M\}$, where $M$ is the number of colluding sets and $\mathcal{T}_m \subseteq [1:N]$, $\forall m \in [1:M]$ is the $m$-th colluding set in $\mathcal{P}$. The representation $\mathcal{P}$ means that the databases in set $\mathcal{T}_m$ may collude, and there are $M$ such colluding sets. As an example, for $N=4$ databases, the $2$-colluding case considered by \cite{sun2018colluding} is denoted as $\mathcal{P}=\{\{1,2\}, \{1,3\}, \{1,4\}, \{2,3\}, \{2,4\}, \{3,4\}\}$, and the disjoint collusion pattern considered in 
\cite{jia2017disjoint} would cover cases such as $\mathcal{P}=\{\{1\},\{2,3\}, \{3,4\}, \{2,4\}\}$, $\mathcal{P}=\{\{1,2\},\{3,4\}\}$ etc. Note that the defined collusion pattern $\mathcal{P}$ satisfy the following two constraints: 
1) we only include the maximal colluding set as elements of $\mathcal{P}$. For example, if $\{1,2,3\} \in \mathcal{P}$, then by definition, $\{1,2\}$ is a colluding set too. But we do not include $\{1,2\}$ in $\mathcal{P}$ for ease of representation; 
2) all databases must appear in at least one element of $\mathcal{P}$, because at the very least, the privacy of the user must be preserved at each single database, which is the requirement of the original PIR problem \cite{sun2017capacity}.

To protect the privacy of the  user, we require that databases that are in a colluding set can not learn anything about the desired message index $\theta$, i.e., 
\begin{align}
(Q_{\mathcal{T}}^{[1]}, A_\mathcal{T}^{[1]}, W_{1:K}) \sim (Q_{\mathcal{T}}^{[\theta]}, A_\mathcal{T}^{[\theta]}, W_{1:K}), \quad  \forall \theta \in [1:K], \quad \forall  \mathcal{T} \in \mathcal{P}.\label{PrivacyConstraint}
\end{align}

The rate of the PIR problem with collusion pattern $\mathcal{P}$, denoted as $R_\mathcal{P}$, is defined as the ratio between the message size $L$ and the total number of downloaded information from the databases,
i.e.,
\begin{align}
R_\mathcal{P}=  \frac{L}{\sum_{n=1}^{N} H(A_{n}^{[\theta]})} \label{DefineR},
\end{align}
which is not a function of $\theta$ due to the privacy constraint in (\ref{PrivacyConstraint}). 
The capacity of the PIR problem with collusion pattern $\mathcal{P}$ is $C_\mathcal{P}=\sup R_{\mathcal{P}}$, where the supremum is over all possible retrieval schemes.

We define an incidence matrix $\mathbf{B}_\mathcal{P}$, of size $N \times M$, to describe the collusion pattern $\mathcal{P}$, where if DB $n$ is in the $m$-th colluding set in $\mathcal{P}$, we let the $(n,m)$-th element of $\mathbf{B}_\mathcal{P}$ be $1$, otherwise, it is zero. For example, $\mathcal{P}=\{\{1,2\},\{2,3\},\{2,4\},\{1,3,4\}\}$ would correspond to an incidence matrix of
\begin{align}
\mathbf{B}_\mathcal{P}=\begin{bmatrix}
1 &0 &0 &1\\
1 & 1 & 1 & 0\\
0 & 1 & 0 & 1\\
0 & 0 & 1 & 1
\end{bmatrix}. \nonumber
\end{align}
Throughout the paper, we will denote the $k \times 1$ column vector of all ones as $\mathbf{1}_k$, and the $k \times 1$ column vector of all zeros as $\mathbf{0}_k$. $\mathbf{I}_k$ is the size $k \times k$ identity matrix and when the size is evident, we write it as $\mathbf{I}$. Similarly, $\mathbf{0}_{k \times i}$ is the size $k \times i$ matrix of all zeros, and when the size is evident, we write it as $\mathbf{0}$. 

\section{Main Results}
The main result of the paper is the PIR capacity under arbitrary collusion patterns for replicated databases, as shown in the next theorem. 
\begin{Theo} \label{main}
The capacity of the PIR problem under collusion pattern $\mathcal{P}$ for replicated databases is
\begin{align}
C_\mathcal{P}=\left(1+\frac{1}{S^*}  +\left(\frac{1}{S^{^*}}\right)^2+\cdots+\left(\frac{1}{S^{*}} \right)^{K-1} \right)^{-1}, \label{CapacityArb}
\end{align}
where $S^*$ is the optimal value of the following linear programming problem, which we will call (LP1),
 \begin{align}
\text{(LP1)} \qquad \max_{\mathbf{y}} \quad & \mathbf{1}_{N}^T \mathbf{y} \nonumber\\
 \text{subject to} \quad & \mathbf{B}^T_\mathcal{P} \mathbf{y} \leq\mathbf{1}_M \label{Constraint02}\\
 & \mathbf{y} \geq \mathbf{0}_N,  \label{Positive01}
 \end{align}
 where $\mathbf{B}_\mathcal{P}$ is the incidence matrix, of size $N \times M$, of the collusion pattern $\mathcal{P}$.
\end{Theo}

Theorem \ref{main} will be proved in the following section. We will first show that (\ref{CapacityArb}) is achievable when the amount of data queried to each database is proportional to the optimal solution $\mathbf{y}^*$ of (LP1). Next, we present a converse theorem where the upper bound on capacity has the same form as (\ref{CapacityArb}) with $S^*$ replaced by $S_2$, and $S_2$ is the optimal value of another linear programming problem (LP2). Finally, we show that (LP1) and (LP2) are dual problems, which means $S^*=S_2$. This concludes the proof that (\ref{CapacityArb}) is the capacity of the PIR problem under arbitrary collusion patterns for replicated databases. 

We make a few remarks here regarding the main result.  
\begin{Remark}
\normalfont Theorem \ref{main} shows that the arbitrary collusion pattern $\mathcal{P}$ affects the capacity of the PIR problem only through the linear programming problem (LP1). More specifically, the capacity formula under arbitrary collusion patterns take on the same form as that of the original PIR problem of (\ref{CapacityPIR}), with $N$ replaced by the optimal solution of (LP1). 
\end{Remark}

\begin{Remark}
\normalfont
Our results coincide with known capacity results of PIR colluding for replicated databases:
\begin{enumerate}
\item In the case of non-colluding databases \cite{sun2017capacity}, the collusion pattern is $\mathcal{P}=\{\{1\}, \{2\}, \cdots, \{N\}\}$, whose incidence matrix is $\mathbf{B}_\mathcal{P}=\mathbf{I}_N$. It is straightforward to see that  the optimal solution to (LP1) is $\mathbf{y}^*=\mathbf{1}_N$, and the corresponding optimal value $S^*=N$. Hence, the capacity formula in (\ref{CapacityArb}) becomes (\ref{CapacityPIR}), consistent with \cite{sun2017capacity}. 
\item In the case of $T$-colluding databases \cite{sun2018colluding}, the collusion pattern $\mathcal{P}$ consists of all size $T$ subsets of $[1:N]$, and there are a total of $N \choose T$ many colluding sets, i.e., $M={N \choose T}$. The corresponding incidence  matrix of size $N \times M$ consists of  $N \choose T$ columns, each with $T$ number of $1$s and $N-T$ number of $0$s. It is straightforward to see that the optimal solution to (LP1) is $\mathbf{y}^*=\frac{1}{T}\mathbf{1}_N$, and the corresponding optimal value $S^*=\frac{N}{T}$. Hence, the capacity formula in (\ref{CapacityArb}) becomes (\ref{CapacityTPIR}), consistent with \cite{sun2018colluding}. 
\item In the case $T$-colluding cyclically contiguous databases \cite[Appendix D]{sun2018private}, the collusion pattern is $\mathcal{P}=\{\{1,2,\cdots,T\}, \{2, 3,\cdots, T+1\}, \cdots, \{N, 1,2,\cdots, T-1\}\}$, where $M=N$. The transpose of the corresponding incidence matrix, i.e., $\mathbf{B}_\mathcal{P}^T$, is a circulant matrix, where the first row consists of $T$ number of 1s followed by $N-T$ number of 0s. It is straightforward to see that though the incidence matrix is different than that of the $T$-colluding case, the optimal solution $\mathbf{y}^*$, and hence the optimal value $S^*$, is the same. Thus, the capacity formula in (\ref{CapacityArb}) becomes (\ref{CapacityTPIR}), consistent with \cite[Appendix D]{sun2018private}. 
\item In the case of disjoint colluding set \cite{jia2017disjoint}, the $N$ servers are split into $J$ disjoint sets, where Set $j$ consists of $N_j$ databases, $j \in [1:J]$. Within Set $j$, up to $T_j$ databases may collude, where $T_j \leq N_j$. The corresponding incidence matrix to this collusion pattern is
\begin{align}
\mathbf{B}_\mathcal{P}=\begin{bmatrix} \mathbf{B}_1 & \mathbf{0} & \cdots & \mathbf{0}\\
 \mathbf{0} & \mathbf{B}_2 & \cdots & \mathbf{0}\\
\vdots & \vdots & \vdots & \vdots\\
 \mathbf{0} &  \mathbf{0} &\cdots & \mathbf{B}_J\end{bmatrix},  \nonumber
\end{align}
where $\mathbf{B}_j$ is an $N_j \times {N_j \choose T_j }$ matrix, with each column consisting of $T_j$ 1s and $N_j-T_j$ 0s, $j \in [1:J]$. It is straightforward to see that the optimal solution to (LP1) is $\mathbf{y}^*=\Big[ \underbrace{\frac{1}{T_1}  \cdots  \frac{1}{T_1}}_{N_1}  \underbrace{\frac{1}{T_2}  \cdots  \frac{1}{T_2}}_{N_2} \cdots \underbrace{\frac{1}{T_J}  \cdots  \frac{1}{T_J}}_{N_J}\Big]^T$.
The corresponding optimal value $S^*=\sum_{j=1}^J \frac{N_j}{T_j}$. Hence, the capacity formula in (\ref{CapacityArb}) becomes $\Bigg(1+ \left(\sum_{j=1}^J \frac{N_j}{T_j} \right)^{-1}+\left(\sum_{j=1}^J \frac{N_j}{T_j} \right)^{-2}+\cdots \break +\left(\sum_{j=1}^J \frac{N_j}{T_j} \right)^{-(K-1)} \Bigg)^{-1}$, consistent with \cite[Theorem 2]{jia2017disjoint}.
\end{enumerate}
\end{Remark}

\section{Proofs}
\subsection{Achievability} \label{SecAch}
Recall that for each collusion pattern $\mathcal{P}$, there is a corresponding incidence matrix $\mathbf{B}_\mathcal{P}$, as defined in Section \ref{secSystemModel}. Let $\mathbf{y}=\begin{bmatrix} y_1 & y_2& \cdots &y_N \end{bmatrix}^T$ be a feasible and rational solution of (LP1), i.e., $\mathbf{y}$ consists of rational elements, and it satisfies the constraints (\ref{Constraint02}) and (\ref{Positive01}). Let the value of the objective function in (LP1) corresponding to $\mathbf{y}$ be $S$, i.e., $S=\sum_{n=1}^N y_n$. Then, we have the following achievability theorem. 
 \begin{Theo} \label{TheoremAchAnyY}
Consider the PIR problem with collusion pattern $\mathcal{P}$, whose incidence matrix is $\mathbf{B}_\mathcal{P}$. Suppose $\mathbf{y}$ is a rational and feasible solution of (LP1) and $S=\mathbf{1}_N^T \mathbf{y}$. Then the following rate is achievable, i.e., 
\begin{align}
C_\mathcal{P} \geq \left(1+\frac{1}{S}  +\left(\frac{1}{S}\right)^2+\cdots+\left(\frac{1}{S} \right)^{K-1} \right)^{-1}. \label{AchRateAnyY}
\end{align}
\end{Theo}
\begin{IEEEproof}
The details of the proof of Theorem \ref{TheoremAchAnyY}, along with an illustrative example, is provided in Appendix \ref{AppAch}. The proof follows very similarly to \cite[Section IV.D]{sun2018colluding}, and we note the difference here: 1) In place of $N^K$ in  \cite[Section IV.D]{sun2018colluding}, we have $L$, which is the message length. $L$ will be chosen such that the number of $k$-sum symbols downloaded from each of the databases is an integer, $k \in [1:K]$.  Such an $L$ can be found since $\mathbf{y}$ is rational. 2) In place of $\frac{N}{T}$ in \cite[Section IV.D]{sun2018colluding}, we have $S$. 3) Rather than distributing the queries evenly among all databases, we distribute the queries among the databases proportionally according to $(\mathbf{y}, S)$, more specifically, the number of queries to Database $n$ is based on the proportion $\frac{y_n}{S}$, $n \in [1:N]$. 
\end{IEEEproof}
\begin{Remark}
\normalfont
The main novelty in our achievable scheme is, rather than distributing the queries evenly among all databases, we propose distributing the queries proportionally according to $(\mathbf{y}, S)$, i.e., the number of queries to Database $n$ is based on the proportion $\frac{y_n}{S}$, $n \in [1:N]$. First of all, this is possible because $\mathbf{y}$ satisfies the constraint in (\ref{Positive01}), which means $y_n \geq 0$, $n \in [1:N]$. Secondly, $\mathbf{y}$ that satisfies the constraint (\ref{Constraint02}) will gurantee the user's privacy. This can be intuitively explained as follows: the databases in each colluding $\mathcal{T}_m \in \mathcal{P}$ can not request too many symbols, i.e., the $m$-th element of $\mathbf{B}_\mathcal{P}^T \mathbf{y}$ is no greater than 1, otherwise, the dependency of the undesired symbols will be revealed to the colluding databases in $\mathcal{T}_m$, violating the privacy of the user.  
\end{Remark}

\begin{Remark}
\normalfont
 It is easy to see that $\mathbf{y}=\frac{1}{N} \mathbf{1}_N$ is a feasible and rational solution. 
 The corresponding $S=\mathbf{1}_N^T \mathbf{y}=1$. This is the suboptimal retrieval scheme of downloading all $K$ messages, evenly from all the databases. 
\end{Remark}

Note that the right-hand side of (\ref{AchRateAnyY}) is an increasing function of $S$. Based on the result of Theorem \ref{TheoremAchAnyY}, to find the largest possible achievable rate, we should find the maximum $S=\sum_{n=1}^N y_n$ achievable over all $\mathbf{y}$ satisfying (\ref{Constraint02}) and (\ref{Positive01}). Applying Theorem \ref{TheoremAchAnyY} for the optimal solution of (LP1), i.e., $(\mathbf{y}^*, S^*)$, and noting that $\mathbf{y}^*$ is rational due to the fact that the objective function and the linear constraints in (LP1) are both with integer coefficients,  the rate of Theorem \ref{main} is achievable.

\subsection{Converse} \label{SecConverse}
Recall that for each collusion pattern $\mathcal{P}$, there is a corresponding incidence matrix $\mathbf{B}_\mathcal{P}$, as defined in Section \ref{secSystemModel}. 
 Consider the following linear programming problem, which will be called (LP2),
 \begin{align}
\text{(LP2)} \qquad \min_{\mathbf{x}} \quad & \mathbf{1}_{M}^T \mathbf{x} \nonumber\\
 \text{subject to} \quad & \mathbf{B}_\mathcal{P} \mathbf{x} \geq \mathbf{1}_N \label{Constraint01}\\
 & \mathbf{x} \geq \mathbf{0}_{M}. \label{Positive02}
 \end{align}
 Let $\mathbf{x}=\begin{bmatrix} x_1 & x_2& \cdots &x_M \end{bmatrix}^T$ be a feasible and rational solution of (LP2), i.e., $\mathbf{x}$ consists of rational elements, and it satisfies the constraints (\ref{Constraint01}) and (\ref{Positive02}). Let the value of the objective function in (LP2) corresponding to $\mathbf{x}$ be $S_2$, i.e., $S_2=\sum_{m=1}^M x_m$.  We have the following converse theorem. 
 \begin{Theo} \label{TheoremConverseAnyX}
Consider the PIR problem with collusion pattern $\mathcal{P}$, whose incidence matrix is $\mathbf{B}_\mathcal{P}$. Suppose $\mathbf{x}$ is a rational and feasible solution of (LP2) and $S_2=\mathbf{1}_M^T \mathbf{x}$. Then, the capacity of the PIR problem is upper bounded by 
\begin{align}
C_\mathcal{P} \leq \left(1+\frac{1}{S_2} +\left(\frac{1}{S_2}\right)^2+\cdots+\left(\frac{1}{S_2}\right)^{K-1} \right)^{-1}. \label{GeneralConverse}
\end{align}
\end{Theo}
\begin{IEEEproof}
The details of the proof is provided in Appendix \ref{proof01}. We comment on the main idea here. 
 Using standard PIR converse techniques such as those in \cite{sun2017capacity}, we can obtain 
for $k=2,3,\cdots,K$,  
\begin{align}
H(A_{1:N}^{[k-1]}|W_{1:k-1}, Q_{1:N}^{[k-1]}) &\geq   H(A_{\mathcal{T}_{m}}^{[k]}|W_{1:k-1}, Q_{1:N}^{[k]}), \quad m=1,2,\cdots,M. \label{idea01}
\end{align}
For each $m \in [1:M]$, multiply both sides of (\ref{idea01}) by $x_m$, which is the $m$-th element of $\mathbf{x}$. Note that $\mathbf{x}$ satisfies (\ref{Positive02}), which means that we are multiplying non-negative numbers and the sign of the inequality does not need to be changed. Then, adding all these $M$ inequalities together, we obtain
\begin{align}
S_2 \cdot H(A_{1:N}^{[k-1]}|W_{1:k-1}, Q_{1:N}^{[k-1]}) &\geq \sum_{m=1}^M x_m  H(A_{\mathcal{T}_{m}}^{[k]}|W_{1:k-1}, Q_{1:N}^{[k]}), \label{Ping02}
\end{align}
where we have used the definition of $S_2$, i.e., $S_2=\sum_{m=1}^M x_m$. 
The fact that $\mathbf{x}$ is rational and non-negative means that there exist non-negative integers $G_\mathbf{x}^1$, $G_\mathbf{x}^2$,$\cdots$, $G_\mathbf{x}^{M}$, $G_{\mathbf{x}}$, such that each $x_m$ can be expressed as $x_m=\frac{G_\mathbf{x}^m}{G_{\mathbf{x}}}$, $m \in [1: M]$. Thus, we have
\begin{align}
&G_{\mathbf{x}} \sum_{m=1}^M x_m  H(A_{\mathcal{T}_{m}}^{[k]}|W_{1:k-1}, Q_{1:N}^{[k]}) 
=  \sum_{m=1}^M G_\mathbf{x}^m  H(A_{\mathcal{T}_{m}}^{[k]}|W_{1:k-1}, Q_{1:N}^{[k]}).  \label{Sub0101}
\end{align}
Since $G_\mathbf{x}^m$, $m \in [1:M]$ are integers, the right-hand side of (\ref{Sub0101}) can be written as a summation of the form 
\begin{align}
\sum_{v=1}^V H(A_{\tilde{\mathcal{T}}_v}^{[k]}|W_{1:k-1}, Q_{1:N}^{[k]}), \label{GeneralSumAppMain}
\end{align} 
where $V$ is a positive integer, and $\tilde{\mathcal{T}}_v \subseteq [1:N]$, for $v \in [1:V]$. 

We have the following results for a summation of the form (\ref{GeneralSumAppMain}):
we say that the summation in (\ref{GeneralSumAppMain})
satisfies the  \emph{even property} with the number $G$, if the number of times $n$ appears in $\mathcal{A} \triangleq \{\tilde{\mathcal{T}}_1, \tilde{\mathcal{T}}_2, \cdots \tilde{\mathcal{T}}_V \}$ is equal to $G$ for each $n \in [1:N]$. For a summation that satisfies the even property, we have 
\begin{align}
\sum_{v=1}^V &H(A_{\tilde{\mathcal{T}}_v}^{[k]}|W_{1:k-1}, Q_{1:N}^{[k]}) \geq G \cdot H(A_{1:N}^{[k]}|W_{1:k-1},Q_{1:N}^{[k]}), \quad k=2,3,\cdots,K, \label{MoPing01}
\end{align}
which follows by applying the sub-modular property of the entropy function multiple times.

In the case of $\mathbf{B}_\mathcal{P} \mathbf{x}=\mathbf{1}_N$, the sum on the right-hand side of (\ref{Sub0101}) satisfies the even property with $G=G_\mathbf{x}$. 
In the case of $\mathbf{B}_\mathcal{P} \mathbf{x}>\mathbf{1}_N$, after writing the right-hand side of (\ref{Sub0101}) in the form of (\ref{GeneralSumAppMain}), we may delete some indices of $n$ in sets $\tilde{\mathcal{T}}_1, \tilde{\mathcal{T}}_2, \cdots \tilde{\mathcal{T}}_V $, 
until each $n$ appears only $G_{\mathbf{x}}$ number of times. This gives us a lower bound to the right-hand side of (\ref{Sub0101}), and this lower bound is a summation that satisfies the even property with the number $G_x$. Hence, for all cases of $\mathbf{B}_\mathcal{P} \mathbf{x} \geq \mathbf{1}_N$, we have
%
\begin{align}
\sum_{m=1}^M G_\mathbf{x}^m  H(A_{\mathcal{T}_{m}}^{[k]}|W_{1:k-1}, Q_{1:N}^{[k]}) \geq G_\mathbf{x}  H(A_{1:N}^{[k]}|W_{1:k-1},Q_{1:N}^{[k]}), \quad k=2,3,\cdots, K. \label{MoPing02}
\end{align}
Utilizing (\ref{Ping02}), (\ref{Sub0101}) and (\ref{MoPing02}), we may obtain the induction argument
\begin{align}
S_2 H(A_{1:N}^{[k-1]}|W_{1:k-1}, Q_{1:N}^{[k-1]}) \geq L+H(A_{1:N}^{[k]}|W_{1:k}, Q_{1:N}^{[k]}), \quad k=2,3,\cdots,K, \nonumber
\end{align}
from which the result of Theorem \ref{TheoremConverseAnyX}
follows from standard PIR converse techniques such as those in \cite{sun2017capacity}. 
\end{IEEEproof}

\begin{Remark}
\normalfont
The main novelty of our converse proof is proving (\ref{MoPing02}), which is used in place of Han's inequality for $T$-colluding \cite{sun2018colluding},  for the induction argument of the converse. We show that when $\mathbf{x}$ satisfies constraints (\ref{Constraint01}) and (\ref{Positive02}), the sum corresponding to $\mathbf{x}$ either satisfies the even property or a lower bound of it satisfies the even property, resulting in (\ref{MoPing02}). 
\end{Remark}

The reason why $\mathbf{x}$ has to satisfy (\ref{Constraint01}), (\ref{Positive02}) and is rational is stated in the proof main idea above. Note that the right-hand side of (\ref{GeneralConverse}) is an increasing function of $S_2$. Based on the result of Theorem \ref{TheoremConverseAnyX}, to find the tightest possible upper bound, we should find the minimum $S_2=\sum_{m=1}^M x_m$ achievable over all $\mathbf{x}$ satisfying (\ref{Constraint01}) and (\ref{Positive02}). Applying Theorem \ref{TheoremConverseAnyX} for the optimal solution of (LP2), i.e., $(\mathbf{x}^*, S_2^*)$, and noting that $\mathbf{x}^*$ is rational due to the fact that the objective function and linear constraints in (LP2) are both with integer coefficients,  we have 
\begin{align}
C_\mathcal{P} \leq \left(1+\frac{1}{S_2^*} +\left(\frac{1}{S_2^*}\right)^2+\cdots+\left(\frac{1}{S_2^*}\right)^{K-1} \right)^{-1}. \nonumber
\end{align}

\subsection{Capacity} \label{SecCapacity}
In Sections \ref{SecAch} and \ref{SecConverse}, we have shown that the capacity lower and upper bounds are related to the optimal solutions of two linear programming problems (LP1) and (LP2), i.e., we have
\begin{align}
 \left(1+\frac{1}{S^*} +\left(\frac{1}{S^{*}}\right)^2+\cdots+\left(\frac{1}{S^{*}}\right)^{K-1} \right)^{-1} \leq C_\mathcal{P} \leq \left(1+\frac{1}{S_2^*} +\left(\frac{1}{S_2^{*}}\right)^2+\cdots+\left(\frac{1}{S_2^{*}}\right)^{K-1} \right)^{-1}, \nonumber
\end{align}
where $S^*$  and $S_2^*$ are the optimal solutions to (LP1) and (LP2), respectively. It is easy to see that (LP1) and (LP2) are actually dual problems of each other, which means $S^*=S_2^*$. Hence, we have found the capacity of the PIR problem under arbitrary collusion pattern $\mathcal{P}$ for replicated databases, as described in Theorem \ref{main}. 

\section{Some Examples}
To aid in a better understanding of the PIR problem under arbitrary collusion patterns for replicated databases, we provide several examples. For ease of understanding, we let $K=2$ messages in all examples. 
\subsection{$N=5$ and $\mathcal{P}_1=\{\{1,2,3\}, \{1,4\}, \{2,4\}, \{3,4\}, \{5\}\}$}
The corresponding incidence matrix is
\begin{align}
\mathbf{B}_{\mathcal{P}_1}=\begin{bmatrix} 1& 1& 0& 0& 0\\
        1& 0 &1& 0 &0 \\
        1 &0 &0 &1 &0\\
        0 &1 &1 &1 &0\\
       0 &0& 0 &0& 1\end{bmatrix}.  \nonumber
\end{align}
The optimal solution to (LP1) is $\mathbf{y}^*=\begin{bmatrix} \frac{1}{3} & \frac{1}{3}& \frac{1}{3}& \frac{2}{3}& 1\end{bmatrix}^T$ and the corresponding optimal value is $\frac{8}{3}$. The optimal solution to (LP2) is $\mathbf{x}^*=\begin{bmatrix} \frac{2}{3} & \frac{1}{3}& \frac{1}{3}& \frac{1}{3}& 1\end{bmatrix}^T$ and the corresponding optimal value is also $\frac{8}{3}$. 

The achievability scheme is as follows: let the message size $L=64$. Further let $\mathbf{U}_1$ and $\mathbf{U}_2 \in \mathbb{F}_q^{64 \times 64}$ be two random matrices chosen privately by the user, independently and uniformly among all $64 \times 64$ full-rank matrices over $\mathbb{F}_q$. Suppose the desired message is $W_1$, then the encoding becomes
\begin{align}
a_{[1:64]}&=\mathbf{U}_1 W_1, \nonumber\\
b_{[1:64]}&=\mathbf{MDS}_{64 \times 24}\mathbf{U}_2 \left[(1:24),: \right] W_2. \nonumber
\end{align}
The query structure is shown in Table \ref{Example0101}. 
\begin{table}[htbp]  
        \caption{Query Table for $N=5$, $K=2$ and Collusion Pattern $\mathcal{P}_1$ corresponding to $\mathbf{y}^*$}
        \begin{center}
            \begin{tabular}{|c|c|c|c|c|} 
                \hline
                DB $1$ & DB $2$ & DB3 & DB4 & DB 5 \\
                \hline
                $a_1, a_2, a_3$ & $a_4, a_5, a_6$ & $a_7, a_8, a_9$ & $a_{10}, a_{11},  a_{12} $ & $a_{16}, a_{17}, a_{18} $\\
                 &  &  & $a_{13}, a_{14},  a_{15}  $ & $a_{19}, a_{20}, a_{21} $\\
                 &  &  &  & $a_{22}, a_{23}, a_{24} $\\
                \hline
                 $b_1, b_2, b_3$ & $b_4, b_5, b_6$ & $b_7, b_8, b_9$ & $b_{10}, b_{11},  b_{12} $ & $b_{16}, b_{17}, b_{18} $\\
                 &  &  & $b_{13}, b_{14},  b_{15}  $ & $b_{19}, b_{20}, b_{21} $\\
                 &  &  &  & $b_{22}, b_{23},b_{24} $\\
                \hline
                 $a_{25}+b_{25} $ &  $a_{30}+b_{30} $  &$a_{35}+b_{35} $ & $a_{40}+b_{40}$  &  $a_{50}+b_{50}$\\
                  $\vdots$ & $\vdots$ &$\vdots$ & $\vdots$&  $\vdots$\\
                 $a_{29}+b_{29} $ &  $a_{34}+b_{34}$  &$a_{39}+b_{39}$ & $a_{44}+b_{44}$&  $a_{54}+b_{54}$\\
                  &   & & $a_{45}+b_{45}$  &  $a_{55}+b_{55}$\\
                 & & & $\vdots$&  $\vdots$\\
                  &   & & $a_{49}+b_{49}$&  $a_{59}+b_{59}$\\
                   &   & &   &  $a_{60}+b_{60}$\\
                 & & & &  $\vdots$\\
                  &   & &&  $a_{64}+b_{64}$\\
                \hline
                          \end{tabular} \label{Example0101}
        \end{center}
\end{table}

The decoding constraint and the achievable rate of $\left( 1+\frac{3}{8}\right)^{-1}$ is simple to check. As for the privacy constraint, for colluding set $\{1,2,3\}$, the 3 databases together see 24 $a$s and 24 $b$s.  Due to the (64, 24) MDS code used, these three databases when colluding sees 24 independent $a$s and 24 independent $b$s, and thus, they can not tell if $a$ or $b$ is the desired message. This holds true for colluding sets $\{1,4\}, \{2,4\}, \{3,4\}$ as well. Each colluding set sees 24 independent $a$s and 24 independent $b$s. As for Database 5, who do not collude with anyone, it sees 24 independent $a$s and 24 independent $b$s by itself. So from this example, we can see that the databases who collude more with others will be queried less, and the databases who collude less with others will get queried more. The heterogeneity of the collusion pattern naturally results in asymmetric database downloading. 

As for the converse, according to $\mathbf{x}^*=\begin{bmatrix} \frac{2}{3} & \frac{1}{3}& \frac{1}{3}& \frac{1}{3}& 1\end{bmatrix}^T$, choose $G_{\mathbf{x}^*}=3$, $G_{\mathbf{x}^*}^1=2$, $G_{\mathbf{x}^*}^2=G_{\mathbf{x}^*}^3=G_{\mathbf{x}^*}^4=1$ and $G_{\mathbf{x}^*}^5=3$. For this example, the proof of the key step (\ref{MoPing02}) is as follows: 
\begin{align}
&2H(A_{\{1,2,3\}}^{[2]}|W_{1}, Q_{1:N}^{[2]})+H(A_{\{1,4\}}^{[2]}|W_{1}, Q_{1:N}^{[2]}) +H(A_{\{2,4\}}^{[2]}|W_{1}, Q_{1:N}^{[2]})\nonumber\\&+H(A_{\{3,4\}}^{[2]}|W_{1}, Q_{1:N}^{[2]}) +3H(A_{\{5\}}^{[2]}|W_{1}, Q_{1:N}^{[2]}) \label{ExampleStep01}\\
\geq & H(A_{\{1,2,3,4\}}^{[2]}|W_{1}, Q_{1:N}^{[2]})+H(A_{\{1,2,3\}}^{[2]}|W_{1}, Q_{1:N}^{[2]})+H(A_{\{1\}}^{[2]}|W_{1}, Q_{1:N}^{[2]}) +H(A_{\{2,4\}}^{[2]}|W_{1}, Q_{1:N}^{[2]})\nonumber\\&+H(A_{\{3,4\}}^{[2]}|W_{1}, Q_{1:N}^{[2]}) +3H(A_{\{5\}}^{[2]}|W_{1}, Q_{1:N}^{[2]}) \nonumber\\
\geq & H(A_{[1:5]}^{[2]}|W_{1}, Q_{1:N}^{[2]})+H(A_{\{1,2,3\}}^{[2]}|W_{1}, Q_{1:N}^{[2]})+H(A_{\{1\}}^{[2]}|W_{1}, Q_{1:N}^{[2]}) +H(A_{\{2,4\}}^{[2]}|W_{1}, Q_{1:N}^{[2]})\nonumber\\&+H(A_{\{3,4\}}^{[2]}|W_{1}, Q_{1:N}^{[2]}) +2H(A_{\{5\}}^{[2]}|W_{1}, Q_{1:N}^{[2]}) \nonumber\\
\geq & H(A_{[1:5]}^{[2]}|W_{1}, Q_{1:N}^{[2]})+H(A_{\{1,2,3,4\}}^{[2]}|W_{1}, Q_{1:N}^{[2]})+H(A_{\{1\}}^{[2]}|W_{1}, Q_{1:N}^{[2]}) +H(A_{\{2\}}^{[2]}|W_{1}, Q_{1:N}^{[2]})\nonumber\\&+H(A_{\{3,4\}}^{[2]}|W_{1}, Q_{1:N}^{[2]}) +2H(A_{\{5\}}^{[2]}|W_{1}, Q_{1:N}^{[2]}) \nonumber\\
\geq & 2H(A_{[1:5]}^{[2]}|W_{1}, Q_{1:N}^{[2]})+H(A_{\{1\}}^{[2]}|W_{1}, Q_{1:N}^{[2]}) +H(A_{\{2\}}^{[2]}|W_{1}, Q_{1:N}^{[2]})\nonumber\\
&+H(A_{\{3,4\}}^{[2]}|W_{1}, Q_{1:N}^{[2]}) +H(A_{\{5\}}^{[2]}|W_{1}, Q_{1:N}^{[2]}) \nonumber\\
 \geq &3H(A_{[1:5]}^{[2]}|W_{1}, Q_{1:N}^{[2]}).  \label{ExampleStep02}
\end{align}
As can be seen, the sum in (\ref{ExampleStep01}) satisfies the even property, and therefore, utilizing the sub-modular property of the entropy function multiple times, will give us (\ref{ExampleStep02}). 

This example is a representation of collusion patterns where the optimal solutions $\mathbf{y}^*$ and $\mathbf{x}^*$ to (LP1) and (LP2) both satisfy the constraints (\ref{Constraint02}) and (\ref{Constraint01}) with equality. The key feature of such collusion patterns are 1) each colluding set of databases are queried with the maximum number of independent bits. 2) the summation of the left-hand side of (\ref{MoPing02}), corresponding to the optimal $\mathbf{x}^*$, satisfies the even property. 

\subsection{$N=5$, $\mathcal{P}_2=\{\{1,3,4\}, \{2,3,4\}, \{1,3,5\}, \{2,3,5\}, \{1,4,5\}, \{2,4,5\}, \{3,4,5\}\}$}

The corresponding incidence matrix is
\begin{align}
\mathbf{B}_{\mathcal{P}_2}=\begin{bmatrix} 1 & 0 & 1 & 0 & 1 & 0 & 0 \\0 & 1 & 0 & 1 &0&1&0\\ 1 &1&1&1&0&0&1\\ 1&1&0&0&1&1&1\\ 0&0&1&1&1&1&1\end{bmatrix}. \nonumber
\end{align}
The optimal solution to (LP1) is $\mathbf{y}^*=\begin{bmatrix} 1 & 1& 0& 0& 0\end{bmatrix}^T$ and the corresponding optimal value is $2$. The optimal solution to (LP2) is $\mathbf{x}^*=\begin{bmatrix} \frac{1}{3} & \frac{1}{3}& \frac{1}{3}& \frac{1}{3}& \frac{1}{3}&\frac{1}{3}&0\end{bmatrix}^T$ and the corresponding optimal value is also $2$. Note here that $\mathbf{B}_{\mathcal{P}_2}^T \mathbf{y}=\mathbf{1}_M$ has a unique non-negative solution $\mathbf{y}_0=\begin{bmatrix} \frac{1}{3} & \frac{1}{3}& \frac{1}{3}& \frac{1}{3}& \frac{1}{3}\end{bmatrix}^T$, yielding a cost function of $\mathbf{1}_N^T \mathbf{y}_0=\frac{5}{3}$, however, it is not the optimal solution to (LP1). 

Consider the following two achievability schemes, the first one corresponds to $\mathbf{y}_0$ and the second one corresponds to $\mathbf{y}^*$. 
The achievability scheme corresponding to $\mathbf{y}_0$ is as follows: let the message size $L=25$. Further let $\mathbf{U}_1$ and $\mathbf{U}_2 \in \mathbb{F}_q^{25 \times 25}$ be two random matrices chosen privately by the user, independently and uniformly among all $25 \times 25$ full-rank matrices over $\mathbb{F}_q$. Suppose the desired message is $W_1$, then the encoding becomes
\begin{align}
a_{[1:25]}&=\mathbf{U}_1 W_1, \nonumber\\
b_{[1:25]}&=\mathbf{MDS}_{25 \times 15}\mathbf{U}_2 \left[(1:15),: \right] W_2. \nonumber
\end{align}
The query structure is shown in Table \ref{Example0202}. 
\begin{table}[htbp]  
        \caption{Query Table for $N=5$, $K=2$ and Collusion Pattern $\mathcal{P}_2$ corresponding to $\mathbf{y}_0$}
        \begin{center}
            \begin{tabular}{|c|c|c|c|c|} 
                \hline
                DB $1$ & DB $2$ & DB3 & DB4 & DB 5 \\
                \hline
                $a_1, a_2, a_3$ & $a_4, a_5, a_6$ & $a_7, a_8, a_9$ & $a_{10}, a_{11},  a_{12} $ & $a_{13}, a_{14}, a_{15} $\\
                \hline
                 $b_1, b_2, b_3$ & $b_4, b_5, b_6$ & $b_7, b_8, b_9$ & $b_{10}, b_{11},  b_{12} $ & $b_{13}, b_{14}, b_{15} $\\
                                \hline
                 $a_{16}+b_{16} $ &  $a_{17}+b_{17} $  &$a_{18}+b_{18} $ & $a_{19}+b_{19}$  &  $a_{20}+b_{20}$\\
                 $a_{21}+b_{21} $ &  $a_{22}+b_{22}$  &$a_{23}+b_{23}$ & $a_{24}+b_{24}$&  $a_{25}+b_{25}$\\
                                \hline
                          \end{tabular} \label{Example0202}
        \end{center}
\end{table}

Each colluding set in $\mathcal{P}_2$ consists of three databases, and they each see 15 independent $a$s and 15 independent $b$s. So this scheme satisfy the feature that each colluding set of databases are queried with the maximum number of independent bits. 

The achievability scheme corresponding to the optimal $\mathbf{y}^*=\begin{bmatrix} 1 & 1& 0& 0& 0\end{bmatrix}^T$, i.e., an optimal query scheme, is to let the message size $L=4$. Further let $\mathbf{U}_1$ and $\mathbf{U}_2 \in \mathbb{F}_q^{4 \times 4}$ be two random matrices chosen privately by the user, independently and uniformly among all $4 \times 4$ full-rank matrices over $\mathbb{F}_q$. Suppose the desired message is $W_1$, then the encoding becomes
\begin{align}
a_{[1:4]}&=\mathbf{U}_1 W_1, \nonumber\\
b_{[1:4]}&=\mathbf{MDS}_{4 \times 2}\mathbf{U}_2 \left[(1:2),: \right] W_2. \nonumber
\end{align}
The query structure is shown in Table \ref{Example0203}, where Databases 3 to 5 will not be queried. When Databases 3 to 5 are not queried, from collusion pattern $\mathcal{P}_2$, we can see that only Databases 1 and 2 are left and they do not collude with each other. So the achievability scheme is just $N=2$ databases with no colluding. 
\begin{table}[htbp]  
        \caption{Query Table for $N=5$, $K=2$ and Collusion Pattern $\mathcal{P}_2$ corresponding to $\mathbf{y}^*$}
        \begin{center}
            \begin{tabular}{|c|c|c|c|c|} 
                \hline
                DB $1$ & DB $2$ & DB3 & DB4 & DB 5 \\
                \hline
                $a_1$ & $a_2$ &  &  & \\
                \hline
                 $b_1$ & $b_2$ &  &  & \\
                                \hline
                 $a_{3}+b_{3} $ &  $a_{4}+b_{4} $  & &   &  \\
                                \hline
                          \end{tabular} \label{Example0203}
        \end{center}
\end{table}
In the above optimal query scheme, since $\mathbf{B}_\mathcal{P}^T \mathbf{y}^*>\mathbf{1}_M$, not all colluding sets get queried the maximum number of independent bits, which is 2. More specifically, colluding set  $\{3,4,5\}$  does not get queried at all. 

Comparing the two achievable schemes, we notice that the first achievable scheme has the feature that each colluding set is queried the maximum number of independent bits, and the second one does not. At first glance, it looks like the first scheme is better as it uses each database to the maximal extent. But in fact, the second one is optimal. Hence, we conclude that the optimal scheme does not necessarily have the following feature: each colluding set is queried with the maximum number of independent bits. 

As for the converse, according to $\mathbf{x}^*=\begin{bmatrix} \frac{1}{3} & \frac{1}{3}& \frac{1}{3}& \frac{1}{3}& \frac{1}{3}&\frac{1}{3}&0\end{bmatrix}^T$, choose $G_{\mathbf{x}^*}=3$, $G_{\mathbf{x}^*}^1=\cdots=G_{\mathbf{x}^*}^6=1$ and $G_{\mathbf{x}^*}^7=0$. For this example, the proof of the key step (\ref{MoPing02}) is as follows: 
\begin{align}
&H(A_{\{1,3,4\}}^{[2]}|W_{1}, Q_{1:N}^{[2]})+H(A_{\{2,3,4\}}^{[2]}|W_{1}, Q_{1:N}^{[2]}) +H(A_{\{1,3,5\}}^{[2]}|W_{1}, Q_{1:N}^{[2]})\nonumber\\&+H(A_{\{2,3,5\}}^{[2]}|W_{1}, Q_{1:N}^{[2]}) +H(A_{\{1,4,5\}}^{[2]}|W_{1}, Q_{1:N}^{[2]})
+H(A_{\{2,4,5\}}^{[2]}|W_{1}, Q_{1:N}^{[2]})  \label{Example2Step01}\\
\geq &H(A_{\{1\}}^{[2]}|W_{1}, Q_{1:N}^{[2]})+H(A_{\{2,3,4\}}^{[2]}|W_{1}, Q_{1:N}^{[2]}) +H(A_{\{1,3\}}^{[2]}|W_{1}, Q_{1:N}^{[2]})\nonumber\\&+H(A_{\{2,3,5\}}^{[2]}|W_{1}, Q_{1:N}^{[2]}) +H(A_{\{1,4,5\}}^{[2]}|W_{1}, Q_{1:N}^{[2]})
+H(A_{\{2,4,5\}}^{[2]}|W_{1}, Q_{1:N}^{[2]})\label{Example2Step03}\\
\geq &H(A_{\{1\}}^{[2]}|W_{1}, Q_{1:N}^{[2]})+H(A_{\{1,2,3,4\}}^{[2]}|W_{1}, Q_{1:N}^{[2]}) +H(A_{\{3\}}^{[2]}|W_{1}, Q_{1:N}^{[2]})\nonumber\\&+H(A_{\{2,3,5\}}^{[2]}|W_{1}, Q_{1:N}^{[2]}) +H(A_{\{1,4,5\}}^{[2]}|W_{1}, Q_{1:N}^{[2]})
+H(A_{\{2,4,5\}}^{[2]}|W_{1}, Q_{1:N}^{[2]}) \nonumber\\
\geq &H(A_{\{1\}}^{[2]}|W_{1}, Q_{1:N}^{[2]})+H(A_{[1:5]}^{[2]}|W_{1}, Q_{1:N}^{[2]}) +H(A_{\{3\}}^{[2]}|W_{1}, Q_{1:N}^{[2]})\nonumber\\&+H(A_{\{2,3\}}^{[2]}|W_{1}, Q_{1:N}^{[2]}) +H(A_{\{1,4,5\}}^{[2]}|W_{1}, Q_{1:N}^{[2]})
+H(A_{\{2,4,5\}}^{[2]}|W_{1}, Q_{1:N}^{[2]})\nonumber\\
\geq &H(A_{\{1\}}^{[2]}|W_{1}, Q_{1:N}^{[2]})+2H(A_{[1:5]}^{[2]}|W_{1}, Q_{1:N}^{[2]}) +H(A_{\{3\}}^{[2]}|W_{1}, Q_{1:N}^{[2]})\nonumber\\&+H(A_{\{2,4,5\}}^{[2]}|W_{1}, Q_{1:N}^{[2]})\nonumber\\
 \geq &3H(A_{[1:5]}^{[2]}|W_{1}, Q_{1:N}^{[2]}).  \label{Example2Step02}
\end{align}
Note that the sum in (\ref{Example2Step01}) does not satisfy the even property, and we need to drop some indices so that each index $n \in [1:5]$ appears $G_{\mathbf{x}^*}$ number of times before we can utilize the sub-modular property of the entropy function. This is why we have the lower bound (\ref{Example2Step03}) where we have dropped indices $3,4$ and $5$ once each. Now, (\ref{Example2Step03}) satisfies the even property, and we may utilize the sub-modular property of the entropy function multiple times to obtain (\ref{Example2Step02}). 

Note the complementary slackness conditions in this example. On one hand, we have $\mathbf{B}_\mathcal{P}^T \mathbf{y}^*=\begin{bmatrix}1 & 1&1 \break &1&1&1&0\end{bmatrix}^T$, which means $\mathbf{x}^*_7=0$. So the colluding set $\{3,4,5\}$ does not appear in the converse, and the converse is derived as if colluding set $\{3,4,5\}$ does not exist, which is still a converse. This intuitively explains why even though the colluding set $\{3,4,5\}$ does not get queried the maximum number of independent bits, Table \ref{Example0202} is still an optimal achievable scheme. On the other hand, we have $\mathbf{B}_\mathcal{P} \mathbf{x}^*=\begin{bmatrix}1 & 1&\frac{4}{3} &\frac{4}{3}&\frac{4}{3} \end{bmatrix}^T$, which means $\mathbf{y}^*_3=\mathbf{y}_4^*=\mathbf{y}_5^*=0$. Hence,  in the optimal achievability scheme, we do not use Databases 3,4 and 5 to transmit anything. This also intuitively explains why we may drop some of the indices of $3,4$ and $5$ in the converse proof of (\ref{Example2Step03}) and have the converse to still be tight. 

The converse proof of not writing $\{3,4,5\}$ in (\ref{Example2Step01}) and dropping some of the indices $3,4$ and $5$ in (\ref{Example2Step03}) means that the converse proof is equivalent to the collusion pattern of $N=5$ databases and $\mathcal{P}_2'=\{\{1\}, \{2,3,4\},\{1,3\}, \{2,3,5\}, \{1,4,5\},  \{2,4,5\}\}=\{\{2,3,4\}, \{1,3\}, \{2,3,5\}, \{1,4,5\}, \break \{2,4,5\}\}$, which is a milder collusion pattern than $\mathcal{P}_2$. The achievability proof of not querying Databases $3,4$ and $5$ means that the achievability proof is equivalent to the collusion pattern $N=2$ databases and $\mathcal{P}_2''=\{\{1\}, \{2\}\}$. So the collusion patterns $\mathcal{P}_2, \mathcal{P}_2', \mathcal{P}_2''$ are in fact equivalent in terms of capacity, which means that for $\mathcal{P}_2$ and $\mathcal{P}_2'$, the collusion is so extensive for Databases $3,4$ and $5$, that we may as well not use them and use Databases $1$ and $2$ only, since these two are not colluding with each other.

\subsection{$N=7$, $\mathcal{P}_3=\{\{1,4\},\{2,5\}, \{1,2,3,6\}, \{3,7\}, \{4,5,6,7\}$}
The corresponding incidence matrix is
\begin{align}
\mathbf{B}_{\mathcal{P}_3}=\begin{bmatrix} 1 & 0 & 1 & 0 & 0\\
0 & 1 &  1 &0&0\\ 
0 &0&1&1&0\\ 
1&0&0&0&1\\ 
0&1&0&0&1\\
0 & 0 & 1& 0 & 1\\
0&0&0&1&1
\end{bmatrix}.  \nonumber
\end{align}
The optimal solution to (LP1) is $\mathbf{y}^*=\begin{bmatrix} \frac{1}{3} & \frac{1}{3}& \frac{1}{3}& \frac{1}{3}& \frac{1}{3} & 0 &\frac{1}{3}\end{bmatrix}^T$ and the corresponding optimal value is $2$. The optimal solution to (LP2) is $\mathbf{x}^*=\begin{bmatrix} 0& 0& 1&0&1\end{bmatrix}^T$ and the corresponding optimal value is also $2$. Note here that $\mathbf{B}_{\mathcal{P}_3} \mathbf{x}=\mathbf{1}_N$ has a unique non-negative solution $\mathbf{x}_0=\begin{bmatrix} \frac{1}{2} &\frac{1}{2}& \frac{1}{2}&\frac{1}{2}& \frac{1}{2}\end{bmatrix}^T$, yielding a cost function of $\mathbf{1}_M^T \mathbf{x}_0=\frac{5}{2}$, however, it is not the optimal solution to (LP2). 

In the following, we derive two converses, the first one is based on $\mathbf{x}_0$ and the second one is based on $\mathbf{x}^*$. 
The converse according to $\mathbf{x}_0=\begin{bmatrix} \frac{1}{2} &\frac{1}{2}& \frac{1}{2}&\frac{1}{2}& \frac{1}{2}\end{bmatrix}^T$ is as follows: choose $G_{\mathbf{x}_0}=2, G_{\mathbf{x}_0}^1=\cdots G_{\mathbf{x}_0}^5=1$, we have the following proof for the key step of (\ref{MoPing02}), 
\begin{align}
&H(A_{\{1,4\}}^{[2]}|W_{1}, Q_{1:N}^{[2]})+H(A_{\{2,5\}}^{[2]}|W_{1}, Q_{1:N}^{[2]}) +H(A_{\{1,2,3,6\}}^{[2]}|W_{1}, Q_{1:N}^{[2]})\nonumber\\&+H(A_{\{3,7\}}^{[2]}|W_{1}, Q_{1:N}^{[2]}) +H(A_{\{4,5,6,7\}}^{[2]}|W_{1}, Q_{1:N}^{[2]}) \label{Example3Step01}\\
\geq &H(A_{\{1,2,4,5\}}^{[2]}|W_{1}, Q_{1:N}^{[2]})+H(A_{\{1,2,3,6\}}^{[2]}|W_{1}, Q_{1:N}^{[2]})+H(A_{\{3,7\}}^{[2]}|W_{1}, Q_{1:N}^{[2]}) \nonumber\\&+H(A_{\{4,5,6,7\}}^{[2]}|W_{1}, Q_{1:N}^{[2]})\nonumber\\
\geq &H(A_{\{1,2,3,4,5,6\}}^{[2]}|W_{1}, Q_{1:N}^{[2]})+H(A_{\{1,2\}}^{[2]}|W_{1}, Q_{1:N}^{[2]})+H(A_{\{3,7\}}^{[2]}|W_{1}, Q_{1:N}^{[2]}) \nonumber\\&+H(A_{\{4,5,6,7\}}^{[2]}|W_{1}, Q_{1:N}^{[2]})\nonumber\\
\geq &H(A_{[1:7]}^{[2]}|W_{1}, Q_{1:N}^{[2]})+H(A_{\{1,2\}}^{[2]}|W_{1}, Q_{1:N}^{[2]})+H(A_{\{3\}}^{[2]}|W_{1}, Q_{1:N}^{[2]}) \nonumber\\&+H(A_{\{4,5,6,7\}}^{[2]}|W_{1}, Q_{1:N}^{[2]})\nonumber\\
\geq &2 H(A_{[1:7]}^{[2]}|W_{1}, Q_{1:N}^{[2]}). \nonumber
\end{align}
Since we have $\mathbf{B}_{\mathcal{P}_3} \mathbf{x}_0=\mathbf{1}_N$, the sum in (\ref{Example3Step01}) satisfies the even property. 

The converse according to $\mathbf{x}^*=\begin{bmatrix} 0& 0& 1&0&1\end{bmatrix}^T$ is as follows: pick $G_{\mathbf{x}^*}=1$, $G_{\mathbf{x}^*}^1=G_{\mathbf{x}^*}^2=G_{\mathbf{x}^*}^4=0$, $G_{\mathbf{x}^*}^3=G_{\mathbf{x}^*}^5=1$, we have the following proof for the key step of (\ref{MoPing02}),  
\begin{align}
&H(A_{\{1,2,3,6\}}^{[2]}|W_{1}, Q_{1:N}^{[2]})+H(A_{\{4,5,6,7\}}^{[2]}|W_{1}, Q_{1:N}^{[2]}) \label{Example3Step02}\\
\geq &H(A_{\{1,2,3\}}^{[2]}|W_{1}, Q_{1:N}^{[2]})+H(A_{\{4,5,6,7\}}^{[2]}|W_{1}, Q_{1:N}^{[2]}) \label{Example3Step03}\\
\geq &H(A_{[1:7]}^{[2]}|W_{1}, Q_{1:N}^{[2]}),  \nonumber
\end{align}
where (\ref{Example3Step02}) does not satisfy the even property, and we drop 6 once to obtain (\ref{Example3Step03}) which satisfies the even property. 

When comparing the two converses, the second one seems looser as it involves a dropping of $6$. However, the second one is in fact tighter as it gives a smaller $S$, which is the sum of the elements of $\mathbf{x}$. This example shows that even if the collusion pattern is such that there exists a sum with the even property, it is not necessarily the tightest converse to use. So the optimal scheme does not necessarily have the following feature: the summation of the left-hand side of (\ref{MoPing02}) satisfies the even property.

The achievability scheme corresponding to $\mathbf{y}^*$ is as follows: let the message size $L=12$. Further let $\mathbf{U}_1$ and $\mathbf{U}_2 \in \mathbb{F}_q^{12 \times 12}$ be two random matrices chosen privately by the user, independently and uniformly among all $12 \times 12$ full-rank matrices over $\mathbb{F}_q$. Suppose the desired message is $W_1$, then the encoding becomes
\begin{align}
a_{[1:12]}&=\mathbf{U}_1 W_1, \nonumber\\
b_{[1:12]}&=\mathbf{MDS}_{12 \times 6}\mathbf{U}_2 \left[(1:6),: \right] W_2. \nonumber
\end{align}
The query structure is shown in Table \ref{Example0303}. 
\begin{table}[htbp]  
        \caption{Query Table for $N=5$, $K=2$ and Collusion Pattern $\mathcal{P}_3$ corresponding to $\mathbf{y}^*$}
        \begin{center}
            \begin{tabular}{|c|c|c|c|c|c|c|} 
                \hline
                DB $1$ & DB $2$ & DB3 & DB4 & DB 5 & DB 6 & DB 7 \\
                \hline
                $a_1$ & $a_2$ & $a_3$ & $a_{4}$ & $a_{5}$ & &$a_{6}$\\
                \hline
                  $b_1$ & $b_2$ & $b_3$ & $b_4$ & $b_5$ & &$b_6$\\
                                \hline
                 $a_{7}+b_{7} $ &  $a_{8}+b_{8} $  &$a_{9}+b_{9} $ & $a_{10}+b_{10}$  &  $a_{11}+b_{11}$ & & $a_{12}+b_{12}$\\
                                \hline
                          \end{tabular} \label{Example0303}
        \end{center}
\end{table}

Colluding sets $\{1,4\}, \{2,5\}, \{3,7\}$ each sees only $4$ independent $a$s and $b$s. Colluding sets $\{1,2,3,6\}$ and $\{4,5,6,7\}$ each sees 6 independent $a$s and $b$s, where Database $6$ does not get queried at all because it colludes with many databases. 

Note the complementary slackness conditions in this example. On one hand, we have $\mathbf{B}_\mathcal{P}^T \mathbf{y}^*=\begin{bmatrix}\frac{2}{3} & \frac{2}{3}&1 &\frac{2}{3}&1\end{bmatrix}^T$, which means $\mathbf{x}^*_1=\mathbf{x}^*_2=\mathbf{x}^*_4=0$. So the colluding sets $\{1,4\}, \{2,5\}, \{3,7\}$  do not appear in the converse, and the converse is derived as if colluding sets $\{1,4\}, \{2,5\}, \{3,7\}$ do not exist, which is still a converse. 
This intuitively explains why even though the colluding sets $\{1,4\}, \{2,5\}, \{3,7\}$ do not get queried the maximum number of independent bits, Table \ref{Example0303} is still an optimal achievable scheme. These colluding sets are not the bottleneck. On the other hand, we have $\mathbf{B}_\mathcal{P} \mathbf{x}^*=\begin{bmatrix}1 & 1&1&1&1&2&1 \end{bmatrix}^T$, which means $\mathbf{y}^*_6=0$. Hence,  in the optimal achievability scheme, we do not use Database 6 to transmit anything. This also intuitively explains why we may drop one of indices $6$ in the converse proof of (\ref{Example3Step03}) and have the converse to still be tight. 

The converse proof of not writing $\{1,4\}, \{2,5\}, \{3,7\}$ in (\ref{Example3Step02}) and dropping one of the indices of $6$ in (\ref{Example3Step03}) means that the converse proof is equivalent to the collusion pattern of $N=7$ databases and $\mathcal{P}_3'=\{\{1,2,3\}, \{4,5,6,7\}\}$, which is a milder collusion pattern than $\mathcal{P}_3$. The achievability proof provided in Table \ref{Example0303} is optimal but not unique. In fact, the optimal scheme can be querying any of the following pairs of databases only: $(1,5)$, $(1,6)$, $(1,7)$, $(2,4)$, $(2,6)$, $(2,7)$, $(3,4)$, $(3,5)$, $(3,6)$. Note that we can not query these pairs of databases: $(1,4), (2,5), (3,7)$ as they may collude.

 \section{Conclusions}
 We have found the capacity of the PIR problem under arbitrary collusion patterns for replicated databases. We first link the achievable PIR rate and its converse to the solutions of two linear programming problems. Then, we show that the two seemingly different linear programming problems have the same optimal value. As a result, the achievable PIR rate and its converse meet, yielding the capacity. The techniques used in this paper can be applied to find  the capacity of other PIR variants under arbitrary patterns, such as symmetric PIR \cite{sun2018capacity2}   under arbitrary collusion patterns \cite{Yao19} and PIR with eavesdropper \cite{wang2018capacity} under arbitrary eavesdropping patterns \cite{Yang19}.


\appendices

\section{Proof of Theorem \ref{TheoremAchAnyY}} \label{AppAch}

We first present the following lemma which is a generalization of \cite[Lemma 1]{sun2018colluding} from square invertible matrices to rectangle matrices with full row rank. This lemma will be used to prove that the proposed achievable scheme satisfies the privacy constraint. 
\begin{Lem} \label{AchPriv}
Suppose $\gamma \leq \beta \leq \alpha$. 
Let $\mathbf{U}_1, \mathbf{U}_2, \cdots, \mathbf{U}_K \in \mathbb{F}_q^{\alpha \times \alpha}$ be $K$ random matrices, drawn independently and uniformly from all $\alpha \times \alpha$ full-rank matrices over $\mathbb{F}_q$. Let $\mathbf{G}_1, \mathbf{G}_2, \cdots, \mathbf{G}_K \in \mathbb{F}_q^{\gamma \times \beta}$ be $K$ matrices of dimension $\gamma \times \beta$ with full row rank. Let $\mathcal{I}_1, \mathcal{I}_2, \cdots, \mathcal{I}_K \in \mathbb{N}^{\beta \times 1}$ be $K$ index vectors, each containing $\beta$ distinct indices from $[1: \alpha]$. Then,
\begin{align}
\left(\mathbf{G}_1 \mathbf{U}_1[\mathcal{I}_1,:],  \mathbf{G}_2 \mathbf{U}_2[\mathcal{I}_2,:], \cdots, \mathbf{G}_K \mathbf{U}_K[\mathcal{I}_K,:]\right) \sim \left(\mathbf{U}_1[(1:\gamma),:], \mathbf{U}_2[(1:\gamma),:],\cdots, \mathbf{U}_K[(1:\gamma),:] \right), \label{UsTedious}
\end{align}
where $\mathbf{U}_k[\mathcal{I}_k,:], k \in [1:K]$ is the $\beta \times \alpha$ matrices comprised of the rows of $\mathbf{U}_k$ with indices in $\mathcal{I}_k$. 
\end{Lem}
\begin{IEEEproof}
We use the results of \cite[Lemma 1]{sun2018colluding} to prove Lemma \ref{AchPriv}. 
Form matrices $\mathbf{G}_1', \cdots, \mathbf{G}_K' \in \mathbb{F}_q^{(\beta-\gamma) \times \beta}$ such that $\begin{bmatrix}\mathbf{G}_k \\ \mathbf{G}_k' \end{bmatrix}$ is a $\beta \times \beta$ square and invertible matrix, for all $k \in [1:K]$. This can be done as $\mathbf{G}_1, \cdots, \mathbf{G}_K$  has full row rank. According to \cite[Lemma 1]{sun2018colluding}, we have
\begin{align}
\left(\begin{bmatrix}\mathbf{G}_1 \\ \mathbf{G}_1' \end{bmatrix} \mathbf{U}_1[\mathcal{I}_1,:],  \cdots, \begin{bmatrix}\mathbf{G}_K \\ \mathbf{G}_K' \end{bmatrix} \mathbf{U}_K[\mathcal{I}_K,:]\right) \sim \left(\mathbf{U}_1[(1:\beta),:],\cdots, \mathbf{U}_K[(1:\beta),:] \right).  \label{SunTedious}
\end{align}
Since we have $\begin{bmatrix} \mathbf{G}_k \\ \mathbf{G}_k' \end{bmatrix} \mathbf{U}_k[\mathcal{I}_k,:]=\begin{bmatrix} \mathbf{G}_k \mathbf{U}_k[\mathcal{I}_k,:] \\ \mathbf{G}_k'  \mathbf{U}_k[\mathcal{I}_k,:] \end{bmatrix}$, $k \in [1:K]$, from (\ref{SunTedious}), we have
\begin{align}
\left(\begin{bmatrix} \mathbf{G}_1 \mathbf{U}_1[\mathcal{I}_1,:] \\ \mathbf{G}_1'  \mathbf{U}_1[\mathcal{I}_1,:] \end{bmatrix},  \cdots, \begin{bmatrix} \mathbf{G}_K \mathbf{U}_K[\mathcal{I}_K,:] \\ \mathbf{G}_K'  \mathbf{U}_K[\mathcal{I}_K,:] \end{bmatrix}\right) 
\sim \left(\begin{bmatrix}\mathbf{U}_1[(1:\gamma),:] \\ \mathbf{U}_1[(\gamma+1:\beta),:] \end{bmatrix}, \cdots, \begin{bmatrix}\mathbf{U}_K[(1:\gamma),:] \\ \mathbf{U}_K[(\gamma+1:\beta),:] \end{bmatrix}\right). \nonumber
\end{align}
Since the above bigger matrices have the same distribution, its sub-matrices have the same distribution too, and thus, (\ref{UsTedious}) follows. 
\end{IEEEproof}

\subsection{An Illustrative Example: $N=5$, $K=3$, $\mathcal{P}_4=\{\{1,2,3\}, \{1,3,4\}, \{2,3,4\}, \{1,2,5\}, \{1,3,5\}, \break \{2,3,5\}, \{4,5\}\}$}

The incidence matrix $\mathbf{B}_{\mathcal{P}_4}$ for the collusion pattern is 
\begin{align}
\mathbf{B}_{\mathcal{P}_4}=\begin{bmatrix} 
1 & 1& 0 & 1 &1&0&0\\
1&0&1&1&0&1&0\\
1&1&1&0&1&1&0\\
0&1&1&0&0&0&1\\
0&0&0&1&1&1&1 \end{bmatrix}. \nonumber
\end{align}
Solving (LP1), we obtain the optimal value $S^*=\frac{7}{4}$, and the optimal $\mathbf{y}^*=\begin{bmatrix} \frac{1}{4} & \frac{1}{4} & \frac{1}{4}& \frac{1}{2}&\frac{1}{2}\end{bmatrix}^T$.  The following achievable scheme is based on $(\mathbf{y}^*, S^*)$, but we only use the fact that $\mathbf{y}^*$ is a rational and feasible solution to (LP1). We do not make use of any of its optimal properties.

Pick a message length $L$ such that the following numbers are integers:
$\frac{L}{S^{*2}}\frac{y_1^*}{S^*}=\frac{L}{S^{*2}}\frac{y_2^*}{S^*}=\frac{L}{S^{*2}}\frac{y_3^*}{S^*}=\frac{16}{343}L$, $\frac{L}{S^{*2}}\frac{y_4^*}{S^*}=\frac{L}{S^{*2}}\frac{y_5^*}{S^*}=\frac{32}{343}L$, $\frac{L}{S^{*2}}(S^*-1)\frac{y_1^*}{S^*}=\frac{L}{S^{*2}}(S^*-1)\frac{y_2^*}{S^*}=\frac{L}{S^{*2}}(S^*-1)\frac{y_3^*}{S^*}=\frac{12}{343}L$, $\frac{L}{S^{*2}}(S^*-1)\frac{y_4^*}{S^*}=\frac{L}{S^{*2}}(S^*-1)\frac{y_5^*}{S^*}=\frac{24}{343}L$, and $\frac{L}{S^{*2}}(S^*-1)^2\frac{y_1^*}{S^*}=\frac{L}{S^{*2}}(S^*-1)^2\frac{y_2^*}{S^*}=\frac{L}{S^{*2}}(S^*-1)^2\frac{y_3^*}{S^*}=\frac{9}{343}L$, $\frac{L}{S^{*2}}(S^*-1)^2\frac{y_4^*}{S^*}=\frac{L}{S^{*2}}(S^*-1)^2\frac{y_5^*}{S^*}=\frac{18}{343}L$. It will be seen that the above expressions are the number of symbols downloaded from each dabatase, and thus, they need to be integers. In this example, we may choose the message length $L$ to be $343$. 

Let $\mathbf{U}_1, \mathbf{U}_2, \mathbf{U}_3 \in \mathbb{F}_q^{L \times L}$ represent random matrices chosen privately by the user, independently and uniformly from all $L \times L$ full-rank matrices over $\mathbb{F}_q$. Suppose $W_1$ is the desired message. 

For the undesired message $W_2$, we perform the following encoding
\begin{align}
\begin{bmatrix}
\begin{matrix}
x_{\mathcal{K}_{1}^{[2]}}^{[2]} \\
x_{\mathcal{K}_{1}^{[2]}\cup\{1\}}^{[2]} \\
\end{matrix} \\ \hdashline[2pt/2pt]
\begin{matrix}
x_{\mathcal{K}_{2}^{[2]}}^{[2]} \\
x_{\mathcal{K}_{2}^{[2]}\cup\{1\}}^{[2]}
\end{matrix} 
\end{bmatrix}
=
\begin{bmatrix}
\begin{matrix}
\mathbf{MDS}_{S^*\alpha_1\times\alpha_1} & \mathbf{0}  \\ \hdashline[2pt/2pt]
\mathbf{0} & \mathbf{MDS}_{S^*\alpha_2\times\alpha_2}  \\ \end{matrix}
\end{bmatrix}
\mathbf{U}_{2}\left[\left(1:\frac{L}{S^*}\right),:\right]W_2, \label{K3}
\end{align}
where $\mathcal{K}_1^{[2]}=\{2\}$ and $\mathcal{K}_2^{[2]}=\{2,3\}$. These two sets are all subsets of $[1:K]$ that contains index $2$ but not $1$.  We choose $\alpha_1$ and $\alpha_2$ in (\ref{K3}) as
\begin{align}
\alpha_1 &\triangleq \left(\frac{1}{S^*} \right)^{K-1} \left(S^*-1 \right)^{|\mathcal{K}_1|-1}L=\frac{16}{49}L=112, \label{AlphaDefine01}\\
\alpha_2& \triangleq \left(\frac{1}{S^*} \right)^{K-1} \left(S^*-1 \right)^{|\mathcal{K}_2|-1}L=\frac{12}{49}L=84. \nonumber
\end{align}
Hence, the MDS codes used above is a $(196,112)$ MDS code and a $(147,84)$ MDS code.  
A similar encoding is performed on the undesired message $W_3$,
\begin{align}
\begin{bmatrix}
\begin{matrix}
x_{\mathcal{K}_{1}^{[3]}}^{[3]} \\
x_{\mathcal{K}_{1}^{[3]}\cup\{1\}}^{[3]} \\
\end{matrix} \\ \hdashline[2pt/2pt]
\begin{matrix}
x_{\mathcal{K}_{2}^{[3]}}^{[3]} \\
x_{\mathcal{K}_{2}^{[3]}\cup\{1\}}^{[3]}
\end{matrix} 
\end{bmatrix}
=
\begin{bmatrix}
\begin{matrix}
\mathbf{MDS}_{S^*\alpha_1\times\alpha_1} & \mathbf{0}  \\ \hdashline[2pt/2pt]
\mathbf{0} & \mathbf{MDS}_{S^*\alpha_2\times\alpha_2}  \\ \end{matrix}
\end{bmatrix}
\mathbf{U}_{3}\left[\left(1:\frac{L}{S^*}\right),:\right]W_3, \label{K33}
\end{align}
where $\mathcal{K}_1^{[3]}=\{3\}$ and $\mathcal{K}_2^{[3]}=\{2,3\}$. Note that the MDS codes used in (\ref{K33}) is the same as that used in (\ref{K3}).

For the desired message $W_1$, we perform the following encoding
\begin{align}
\begin{bmatrix}
x_{\mathcal{L}_{1}}^{[1]} \\
x_{\mathcal{L}_{2}}^{[1]} \\
x_{\mathcal{L}_{3}}^{[1]} \\
x_{\mathcal{L}_{4}}^{[1]} 
\end{bmatrix}
= \mathbf{U}_{1}W_{1}, \label{K3Desired}
\end{align}
where $\mathcal{L}_1=\{1\}$, $\mathcal{L}_2=\{1,2\}$, $\mathcal{L}_3=\{1,3\}$ and $\mathcal{L}_4=\{1,2,3\}$. These sets are all the subsets of $[1:K]$ that contain 1. 

In (\ref{K3}), (\ref{K33}) and (\ref{K3Desired}), $x_{\mathcal{K}}^{[k]}$, $\mathcal{K} \subseteq [1:K], k \in [1:K]$ is a column vector with length $\left(\frac{1}{S^*} \right)^{K-1} \left(S^*-1 \right)^{|\mathcal{K}|-1}L$. With the above definitions, it is straightforward to check that the dimensions of the left-hand side is equal to that of the right-hand side in (\ref{K3}), (\ref{K33}) and (\ref{K3Desired}). For this example, we have $x_{\{1\}}^{[1]}$, $x_{\{2\}}^{[2]}$ and $x_{\{3\}}^{[3]}$ are all column vectors with length 
\begin{align}
\beta_1 \triangleq \left(\frac{1}{S^*} \right)^{K-1} L=\frac{16}{49}L=112,  \label{BetaDefine01}
\end{align} 
and $x_{\{1,2\}}^{[1]}$, $x_{\{1,2\}}^{[2]}$,  $x_{\{1,3\}}^{[1]}$, $x_{\{1,3\}}^{[3]}$, $x_{\{2,3\}}^{[2]}$, $x_{\{2,3\}}^{[3]}$ are all column vectors with length 
\begin{align}
\beta_2 \triangleq\left(\frac{1}{S^*} \right)^{K-1} \left(S^*-1 \right)L=\frac{12}{49}L=84, \label{BetaDefine02}
\end{align} 
and $x_{\{1,2,3\}}^{[1]}$, $x_{\{1,2,3\}}^{[2]}$,  $x_{\{1,2,3\}}^{[3]}$ are all column vectors with length 
\begin{align}
\beta_3 \triangleq\left(\frac{1}{S^*} \right)^{K-1} \left(S^*-1 \right)^2L=\frac{9}{49}L=63. \nonumber
\end{align} 
Note that $\beta_i=\alpha_i$, $i=1,2$.

For each $\mathcal{K}=\{1\}, \{2\}, \{3\},\{1,2\}, \{1,3\}, \{2,3\},\{1,2,3\}$, generate the query vector
\begin{align}
\sum_{k \in \mathcal{K}} x_{\mathcal{K}}^{[k]}, \label{GeneralAchK2}
\end{align}
which is a column vector with length $\left(\frac{1}{S^*} \right)^{K-1} \left(S^*-1 \right)^{|\mathcal{K}|-1}L$. We will distribute these elements to the databases according to $\mathbf{y}^*$, which means that a proportion $\frac{y_n^*}{S^*}$ of $\left(\frac{1}{S^*} \right)^{K-1} \left(S^*-1 \right)^{|\mathcal{K}|-1}L$ many queries of (\ref{GeneralAchK2}) is from DB $n$, $n \in [1:N]$ for each $\mathcal{K}=\{1\}, \{2\}, \{3\},\{1,2\}, \{1,3\}, \{2,3\},\break \{1,2,3\}$. More specifically, if we write out the query table, it would be as in Table \ref{K3QueryNan}.
\begin{table}[htbp]  
        \caption{Query Table for $N=5$, $K=3$ and Collusion Pattern $\mathcal{P}_4$ corresponding to $\mathbf{y}^*$}
        \begin{center}
            \begin{tabular}{|c|c|c|c|c|} 
                \hline
                DB $1$ & DB $2$ & DB3 & DB4 & DB 5 \\
                $\frac{y_1^*}{S^*}=\frac{1}{7}$ & $\frac{y_2^*}{S^*}=\frac{1}{7}$ & $\frac{y_3^*}{S^*}=\frac{1}{7}$ & $\frac{y_4^*}{S^*}=\frac{2}{7}$ & $\frac{y_5^*}{S^*}=\frac{2}{7}$\\
                \hline
                $x_{\{1\}}^{[1]}(1:16)$ & $x_{\{1\}}^{[1]}(17:32)$ & $x_{\{1\}}^{[1]}(33:48)$ & $x_{\{1\}}^{[1]}(49:80)$ & $x_{\{1\}}^{[1]}(81:112)$\\
                \hline
                 $x_{\{2\}}^{[2]}(1:16)$ & $x_{\{2\}}^{[2]}(17:32)$ & $x_{\{2\}}^{[2]}(33:48)$ & $x_{\{2\}}^{[2]}(49:80)$ & $x_{\{2\}}^{[2]}(81:112)$\\
                \hline
                 $x_{\{3\}}^{[3]}(1:16)$ & $x_{\{3\}}^{[3]}(17:32)$ & $x_{\{3\}}^{[3]}(33:48)$ & $x_{\{3\}}^{[3]}(49:80)$ & $x_{\{3\}}^{[3]}(81:112)$\\
                \hline
                 $x_{\{1,2\}}^{[1]}(1:12)$ & $x_{\{1,2\}}^{[1]}(13:24)$ & $x_{\{1,2\}}^{[1]}(25:36)$ & $x_{\{1,2\}}^{[1]}(37:60)$ &$x_{\{1,2\}}^{[1]}(61:84)$\\
                 $+x_{\{1,2\}}^{[2]}(1:12)$ & $+x_{\{1,2\}}^{[2]}(13:24)$ & $+x_{\{1,2\}}^{[2]}(25:36)$ & $+x_{\{1,2\}}^{[2]}(37:60)$& $+x_{\{1,2\}}^{[2]}(61:84)$\\
                \hline
                 $x_{\{1,3\}}^{[1]}(1:12)$ & $x_{\{1,3\}}^{[1]}(13:24)$ & $x_{\{1,3\}}^{[1]}(25:36)$ & $x_{\{1,3\}}^{[1]}(37:60)$ &$x_{\{1,3\}}^{[1]}(61:84)$\\
                 $+x_{\{1,3\}}^{[3]}(1:12)$ & $+x_{\{1,3\}}^{[3]}(13:24)$ & $+x_{\{1,3\}}^{[3]}(25:36)$ & $+x_{\{1,3\}}^{[3]}(37:60)$& $+x_{\{1,3\}}^{[3]}(61:84)$\\
                \hline
                 $x_{\{2,3\}}^{[2]}(1:12)$ & $x_{\{2,3\}}^{[2]}(13:24)$ & $x_{\{2,3\}}^{[2]}(25:36)$ & $x_{\{2,3\}}^{[2]}(37:60)$ &$x_{\{2,3\}}^{[2]}(61:84)$\\
                 $+x_{\{2,3\}}^{[3]}(1:12)$ & $+x_{\{2,3\}}^{[3]}(13:24)$ & $+x_{\{2,3\}}^{[3]}(25:36)$ & $+x_{\{2,3\}}^{[3]}(37:60)$& $+x_{\{2,3\}}^{[3]}(61:84)$\\
                \hline
                 $x_{\{1,2,3\}}^{[1]}(1:9)$ & $x_{\{1,2,3\}}^{[1]}(10:18)$ & $x_{\{1,2,3\}}^{[1]}(19:27)$ & $x_{\{1,2,3\}}^{[1]}(28:45)$ &$x_{\{1,2,3\}}^{[1]}(46:63)$\\
                 $+x_{\{1,2,3\}}^{[2]}(1:9)$ & $+x_{\{1,2,3\}}^{[2]}(10:18)$ & $+x_{\{1,2,3\}}^{[2]}(19:27)$ & $+x_{\{1,2,3\}}^{[2]}(28:45)$& $+x_{\{1,2,3\}}^{[2]}(46:63)$\\
                 $+x_{\{1,2,3\}}^{[3]}(1:9)$ & $+x_{\{1,2,3\}}^{[3]}(10:18)$ & $+x_{\{1,2,3\}}^{[3]}(19:27)$ & $+x_{\{1,2,3\}}^{[3]}(28:45)$& $+x_{\{1,2,3\}}^{[3]}(46:63)$\\
                \hline
            \end{tabular} \label{K3QueryNan}
        \end{center}
\end{table}

Now, we check that the decoding constraint is satisfied. Recall that the undesired message $W_2$ is encoded as (\ref{K3}), hence, upon receiving $x_{\{2\}}^{[2]}(1:112)$, the user may calculate $x_{\{1,2\}}^{[2]}(1:84)$ according to the $(196,112)$ MDS code used. Similarly, for undesired message $W_3$, upon receiving $x_{\{3\}}^{[3]}(1:112)$, the user may calculate $x_{\{1,3\}}^{[3]}(1:84)$. Furthermore, upon receiving $x_{\{2,3\}}^{[2]}(1:84)+x_{\{2,3\}}^{[3]}(1:84)$, due to the same $(147,84)$ MDS code used in (\ref{K3}) and (\ref{K33}), the  user may calculate $x_{\{1,2,3\}}^{[2]}(1:63)+x_{\{1,2,3\}}^{[3]}(1:63)$.
Subtracting all the undesired symbols $x^{[2]}$ and $x^{[3]}$, we obtain $x_{\{1\}}^{[1]}(1:112)$, $x_{\{1,2\}}^{[1]}(1:84)$, $x_{\{1,3\}}^{[1]}(1:84)$ and $x_{\{1,2,3\}}^{[1]}(1:63)$ and calculate the desired message $W_1$ according to (\ref{K3Desired}). Thus, by downloading $112 \times 3+84 \times 3+63$ many symbols, we obtain $112+84\times 2+63$ desired symbols, achieving a rate of $\frac{49}{93}$, which is equal to $\left(1+\frac{1}{S^*}+ \left(\frac{1}{S^*} \right)^2\right)^{-1}$.

Now, we check that the privacy constraint is satisfied. 
Recall that $\mathcal{P}_4=\{\{1,2,3\}, \{1,3,4\},  \break \{2,3,4\}, \{1,2,5\}, \{1,3,5\},  \{2,3,5\}, \{4,5\}\}$. Define the set of indices of $x_\mathcal{K}^{[k]}$ retrieved from colluding set $\mathcal{T}_m$ as $\mathcal{I}_\mathcal{K}^{[k]m}$. Then, for the $m$-th colluding set $\mathcal{T}_m \in \mathcal{P}_4$, $m \in [1:M]$, the number of $(x_{\{2\}}^{[2]}, x_{\{1,2\}}^{[2]})$ retrieved from the databases in $\mathcal{T}_m$, i.e., $\left|\mathcal{I}_{\{2\}}^{[2]m} \right| +\left|\mathcal{I}_{\{1,2\}}^{[2]m} \right|$, and the number of $(x_{\{3\}}^{[3]}, x_{\{1,3\}}^{[3]})$ retrieved from the databases in $\mathcal{T}_m$, i.e., $\left|\mathcal{I}_{\{3\}}^{[3]m} \right| +\left|\mathcal{I}_{\{1,3\}}^{[3]m} \right|$, satisfy
\begin{align}
\left|\mathcal{I}_{\{2\}}^{[2]m} \right| +\left|\mathcal{I}_{\{1,2\}}^{[2]m} \right|&=\left|\mathcal{I}_{\{3\}}^{[3]m} \right| +\left|\mathcal{I}_{\{1,3\}}^{[3]m} \right| \nonumber\\
&=\sum_{n \in \mathcal{T}_m} (\beta_1+\beta_2)\frac{y_n^*}{S^*} \nonumber\\
&=\frac{\beta_1+\beta_2}{S^*} \left(\mathbf{B}_{\mathcal{P}_4}^T \mathbf{y}^* \right)_m \label{Incidence01}\\
& \leq \frac{\beta_1+\beta_2}{S^*} \label{LinkLP1} \\
&=\left(\frac{1}{S^*}\right)^{K-1} L \label{BetaAlpha}\\
&=\alpha_1, \label{BetaAlpha01}
\end{align}
where in (\ref{Incidence01}), $(\mathbf{x})_m$ denote the $m$-th element of the vector $\mathbf{x}$, (\ref{LinkLP1}) follows because $\mathbf{y}^*$ satisfies the condition (\ref{Constraint02}) in (LP1), and (\ref{BetaAlpha}), (\ref{BetaAlpha01}) follows from our definition of $\beta$s and $\alpha$s in (\ref{AlphaDefine01}), (\ref{BetaDefine01}) and (\ref{BetaDefine02}).  The derivation from (\ref{Incidence01}) to (\ref{LinkLP1}) clearly shows why in the linear programming (LP1), we have the constraint (\ref{Constraint02}). 
Since the number of $(x_{\{2\}}^{[2]}, x_{\{1,2\}}^{[2]})$ retrieved from $\mathcal{T}_m$ is less than $\alpha_1$, $\mathbf{MDS}_{S^*\alpha_1\times\alpha_1} \left[\mathcal{I}_{\{2\}}^{[2]m} \bigcup \mathcal{I}_{\{1,2\}}^{[2]m},: \right]$ is full row rank. Similarly, $\mathbf{MDS}_{S^*\alpha_1\times\alpha_1} \left[\mathcal{I}_{\{3\}}^{[3]m} \bigcup \mathcal{I}_{\{1,3\}}^{[3]m},: \right]$ is full row rank.

By a similar argument, the number of $(x_{\{2,3\}}^{[2]}, x_{\{1,2,3\}}^{[2]})$ retrieved from the databases in $\mathcal{T}_m$, i.e., $\left|\mathcal{I}_{\{2,3\}}^{[2]m} \right| +\left|\mathcal{I}_{\{1,2,3\}}^{[2]m} \right|$, and the number of $(x_{\{2,3\}}^{[3]}, x_{\{1,2,3\}}^{[3]})$, i.e., $\left|\mathcal{I}_{\{2,3\}}^{[3]m} \right| +\left|\mathcal{I}_{\{1,2,3\}}^{[3]m} \right|$ satisfy
\begin{align}
&\left|\mathcal{I}_{\{2,3\}}^{[2]m} \right| +\left|\mathcal{I}_{\{1,2,3\}}^{[2]m} \right|=\left|\mathcal{I}_{\{2,3\}}^{[3]m} \right| +\left|\mathcal{I}_{\{1,2,3\}}^{[3]m} \right|\nonumber\\
=&\sum_{n \in \mathcal{T}_m} (\beta_2+\beta_3)\frac{y_n^*}{S^*}=\frac{\beta_2+\beta_3}{S^*} \left(\mathbf{B}_{\mathcal{P}_4}^T \mathbf{y}^* \right)_m \leq \frac{\beta_2+\beta_3}{S^*}=\left(\frac{1}{S^*}\right)^{K-1} (S^*-1) L=\alpha_2. \nonumber
\end{align}
Thus, matrices $\mathbf{MDS}_{S^*\alpha_2\times\alpha_2} \left[\mathcal{I}_{\{2,3\}}^{[2]m} \bigcup \mathcal{I}_{\{1,2,3\}}^{[2]m},: \right]$ and $\mathbf{MDS}_{S^*\alpha_2\times\alpha_2} \left[\mathcal{I}_{\{2,3\}}^{[3]m} \bigcup \mathcal{I}_{\{1,2,3\}}^{[3]m},: \right]$ are both of full row rank.

Hence, the matrix 
\begin{align}
\mathbf{G}_2^m \triangleq
\begin{bmatrix}
\begin{matrix}
\mathbf{MDS}_{S^*\alpha_1\times\alpha_1} \left[\mathcal{I}_{\{2\}}^{[2]m} \bigcup \mathcal{I}_{\{1,2\}}^{[2]m},: \right] & \mathbf{0}  \\ \hdashline[2pt/2pt]
\mathbf{0} & \mathbf{MDS}_{S^*\alpha_2\times\alpha_2} \left[\mathcal{I}_{\{2,3\}}^{[2]m} \bigcup \mathcal{I}_{\{1,2,3\}}^{[2]m},: \right]  \\ \end{matrix}
\end{bmatrix} \nonumber
\end{align}
is full row rank, and so is 
\begin{align}
\mathbf{G}_3^m \triangleq
\begin{bmatrix}
\begin{matrix}
\mathbf{MDS}_{S^*\alpha_1\times\alpha_1} \left[\mathcal{I}_{\{3\}}^{[3]m} \bigcup \mathcal{I}_{\{1,3\}}^{[3]m},: \right] & \mathbf{0}  \\ \hdashline[2pt/2pt]
\mathbf{0} & \mathbf{MDS}_{S^*\alpha_2\times\alpha_2} \left[\mathcal{I}_{\{2,3\}}^{[3]m} \bigcup \mathcal{I}_{\{1,2,3\}}^{[3]m},: \right] \\ \end{matrix}
\end{bmatrix}. \nonumber
\end{align}
For notational convenience, let $\mathcal{I}^{[1]m}=\mathcal{I}_{\{1\}}^{[1]m} \bigcup \mathcal{I}_{\{1,2\}}^{[1]m} \bigcup \mathcal{I}_{\{1,3\}}^{[1]m} \bigcup \mathcal{I}_{\{1,2,3\}}^{[1]m}$, which is the indices of $x^{[1]}$ received by databases in $\mathcal{T}_m$. Similarly, define $\mathcal{I}^{[2]m}=\mathcal{I}_{\{2\}}^{[2]m} \bigcup \mathcal{I}_{\{1,2\}}^{[2]m} \bigcup \mathcal{I}_{\{2,3\}}^{[2]m} \bigcup \mathcal{I}_{\{1,2,3\}}^{[2]m}$ and $\mathcal{I}^{[3]m}=\mathcal{I}_{\{3\}}^{[3]m} \bigcup \mathcal{I}_{\{1,3\}}^{[3]m} \bigcup \mathcal{I}_{\{2,3\}}^{[3]m} \bigcup \mathcal{I}_{\{1,2,3\}}^{[3]m}$. Note that $ \frac{L}{S^*}=\alpha_1+\alpha_2 \geq \left|\mathcal{I}^{[1]m} \right|=\left|\mathcal{I}^{[2]m} \right|=\left|\mathcal{I}^{[3]m} \right| \triangleq \tau_m$.
Databases in the colluding set $\mathcal{T}_m$ sees $\Big(x^{[1]}_{\mathcal{I}^{[1]m}} ,x^{[2]}_{\mathcal{I}^{[2]m}},x^{[3]}_{\mathcal{I}^{[3]m}} \Big)$ with distribution $\Big(\mathbf{U}_1\left[\mathcal{I}^{[1]m},:\right] W_1, \mathbf{G}_2^m \mathbf{U}_{2}\left[\left(1:\frac{L}{S^*}\right),:\right] W_2,  \mathbf{G}_3^m \mathbf{U}_{3}\left[\left(1:\frac{L}{S^*}\right),:\right] W_3\Big)$. To use Lemma \ref{AchPriv}, rewrite $\mathbf{U}_1\left[\mathcal{I}^{[1]m},: \right]  =\begin{bmatrix} \mathbf{I}_{\tau_m} & \mathbf{0}_{\tau_m \times (\frac{L}{S^*}-\tau_m)} \end{bmatrix} \begin{bmatrix}\mathbf{U}_1\left[\mathcal{I}^{[1]m},:\right]\\
\mathbf{U}_1\left[ \mathcal{I}_c^{[1]m},: \right] \end{bmatrix}$, where $\mathcal{I}_c^{[1]m}$ is chosen as $\frac{L}{S^*}-\tau_m$ number of indices in $[1:L]$ who are not in $\mathcal{I}^{[1]m}$. Applying Lemma \ref{AchPriv}, we have
\begin{align}
&\Big(\mathbf{U}_1\left[\mathcal{I}^{[1]m},:\right], \mathbf{G}_2^m\mathbf{U}_{2}\left[\left(1:\frac{L}{S^*}\right),:\right], \mathbf{G}_3^m \mathbf{U}_{3}\left[\left(1:\frac{L}{S^*}\right),:\right] \Big) \nonumber\\
& \sim  \left(\mathbf{U}_1[(1:\tau_m),:], \mathbf{U}_2[(1:\tau_m),:],\mathbf{U}_3[(1:\tau_m),:] \right), \quad \forall m \in [1:M], \nonumber
\end{align}
which proves that the retrieval scheme is private.

\subsection{General Achievability Scheme for arbitrary number of messages $K$, arbitrary number of databases $N$ and arbitrary collusion pattern $\mathcal{P}$}

 Let $\mathbf{y}=\begin{bmatrix} y_1 & y_2& \cdots &y_N \end{bmatrix}^T$ be a feasible and rational solution of (LP1), i.e., $\mathbf{y}$ consists of rational elements, and it satisfies the constraints (\ref{Constraint02}) and (\ref{Positive01}). Let the value of the objective function in (LP1) corresponding to $\mathbf{y}$ be $S$, i.e., $S=\sum_{n=1}^N y_n$. 
 
 The encoding of the messages follows the scheme in \cite[Section IV.D]{sun2018colluding} closely with $N^K$ replaced by $L$, the message size, and $\frac{N}{T}$ replaced with $S$. For completeness, we state the scheme here. 

Pick message length $L$ such that the following numbers are integers:
$\left(\frac{1}{S} \right)^{K} \left(S-1 \right)^{k-1}Ly_n$, $k \in [1:K]$, $n\in [1:N]$. 
Note that the above involves $KN$ numbers which is finite. Such an $L$ can be found because $S$ and $y_n$, $n \in [1:N]$ are rational numbers.

Let each message contain $L$ number of symbols from $\mathbb{F}^q$. Let $\mathbf{U}_1, \mathbf{U}_2, \cdots, \mathbf{U}_K \in \mathbb{F}_q^{L \times L}$ represent random matrices chosen privately by the user, independently and uniformly from all $L \times L$ full-rank matrices over $\mathbb{F}_q$. Suppose $W_\theta, \theta \in [1:K]$ is the desired message. 

For each undesired message $k \in [1:K] \setminus \{\theta \}$, we perform the following encoding
\begin{align}
\begin{bmatrix}
\begin{matrix}
x_{\mathcal{K}_{1}^{[k]}}^{[k]} \\
x_{\mathcal{K}_{1}^{[k]}\cup\{\theta\}}^{[k]} \\
\end{matrix} \\ \hdashline[2pt/2pt]
\begin{matrix}
x_{\mathcal{K}_{2}^{[k]}}^{[k]} \\
x_{\mathcal{K}_{2}^{[k]}\cup\{\theta\}}^{[k]}
\end{matrix} \\ \hdashline[2pt/2pt]
\begin{matrix}
\vdots
\end{matrix} \\ \hdashline[2pt/2pt]
\begin{matrix}
x_{\mathcal{K}_{\Delta}^{[k]}}^{[k]} \\
x_{\mathcal{K}_{\Delta}^{[k]}\cup\{\theta\}}^{[k]}
\end{matrix}
\end{bmatrix}
=
\begin{bmatrix}
\begin{matrix}
\mathbf{MDS}_{S\alpha_1\times\alpha_1} & \mathbf{0} & \mathbf{0} & \mathbf{0} \\ \hdashline[2pt/2pt]
\mathbf{0} & \mathbf{MDS}_{S\alpha_2\times\alpha_2} & \mathbf{0} & \mathbf{0} \\ \hdashline[2pt/2pt]
\mathbf{0} & \cdots & \ddots & \mathbf{0} \\ \hdashline[2pt/2pt]
\mathbf{0} & \mathbf{0} & \mathbf{0} & \mathbf{MDS}_{S \alpha_{\Delta}\times\alpha_{\Delta}}
\end{matrix}
\end{bmatrix}
\mathbf{U}_{k}\left[\left(1:\frac{L}{S}\right),:\right]W_k, \label{LongEquation}
\end{align}
where $\mathcal{K}_1^{[k]}, \mathcal{K}_2^{[k]}, \cdots, \mathcal{K}_\Delta^{[k]}$ are the distinct labels we assign to all distinct $\Delta=2^{K-2}$ subsets of $[1:K]$ that contain $k$ and do not contain $\theta$, and $\alpha_i, i \in [1:\Delta]$ is defined as $\left(\frac{1}{S} \right)^{K-1} \left(S-1 \right)^{|\mathcal{K}_i|-1}L$.

For the desired message index $\theta$, we perform the following encoding
\begin{align}
\begin{bmatrix}
x_{\mathcal{L}_{1}^{[\theta]}}^{[\theta]} \\
x_{\mathcal{L}_{2}^{[\theta]}}^{[\theta]} \\
\vdots \\
x_{\mathcal{L}_{\delta}^{[\theta]}}^{[\theta]} \\
\end{bmatrix}
= \mathbf{U}_{\theta}W_{\theta}, \label{Nan06}
\end{align}
where $\mathcal{L}_i^{[\theta]}$ is the distinct labels of the $\delta=2^{K-1}$ subsets of $[1:K]$ that contain $\theta$.

In (\ref{LongEquation}) and (\ref{Nan06}), $x_{\mathcal{K}}^{[k]}$, $\mathcal{K} \subseteq [1:K], k \in [1:K]$ is a column vector with length $\left(\frac{1}{S} \right)^{K-1} \break \left(S-1 \right)^{|\mathcal{K}|-1}L$. With the above definitions, it is straightforward to check that the dimensions of the left-hand side is equal to that of the right-hand side in (\ref{LongEquation}) and (\ref{Nan06}).

For each non-empty subset $\mathcal{K} \subseteq [1:K]$, generate the query vector
\begin{align}
\sum_{k \in \mathcal{K}} x_{\mathcal{K}}^{[k]}, \label{GeneralAch}
\end{align}
which is a column vector with length $\left(\frac{1}{S} \right)^{K-1} \left(S-1 \right)^{|\mathcal{K}|-1}L$. Up until now, our achievable scheme follows the scheme in \cite[Section IV.D]{sun2018colluding} closely with $N^K$ replaced by $L$, the message size, and $\frac{N}{T}$ replaced with $S$.

Rather than distributing the queries evenly among all databases as in \cite[Section IV.D]{sun2018colluding}, here we will distribute these elements to the databases according to $\mathbf{y}$, which means a proportion $\frac{y_n}{S}$ of $\left(\frac{1}{S} \right)^{K-1} \left(S-1 \right)^{|\mathcal{K}|-1}L$ many queries of (\ref{GeneralAch}) is from DB $n$, $n \in [1:N]$, for each $\mathcal{K} \subseteq [1:K]
$. Note that this is possible because $\mathbf{y}$ satisfies the constraint in (\ref{Positive01}), i.e., $y_n \geq 0$, $n \in [1:N]$. 

The decoding constraint is satisfied, following the same proof as \cite[Section IV.D]{sun2018colluding}, with $N^K$ replaced by $L$ and $\frac{N}{T}$ replaced by $S$. So it will not be repeated here. The achievable rate of (\ref{AchRateAnyY}) also follows by replacing $(N^K, \frac{N}{T})$ with $(L,S)$ in \cite[eqn. (32)-(35)]{sun2018colluding}.

Finally, we check that the privacy constraint is satisfied. Define the set of indices of $x_\mathcal{K}^{[k]}$ retrieved from colluding set $\mathcal{T}_m$ as $\mathcal{I}_\mathcal{K}^{[k]m}$. Then, for the $m$-th colluding set $\mathcal{T}_m \in \mathcal{P}$, $m \in [1:M]$, $k \in [1:K] \setminus \{\theta\}$, and $\mathcal{K}_i^{[k]}$, $i \in [1:\Delta]$, %
the number of $(x_{\mathcal{K}_i^{[k]}}^{[k]}, x_{\mathcal{K}_i^{[k]} \bigcup \{\theta\}}^{[k]})$ retrieved from the databases in $\mathcal{T}_m$ satisfy
\begin{align}
&\left|\mathcal{I}_{\mathcal{K}_i^{[k]}}^{[k]m} \right| +\left|\mathcal{I}_{\mathcal{K}_i^{[k]} \bigcup \{\theta\}}^{[k]m} \right| \nonumber\\
&=\sum_{n \in \mathcal{T}_m} \left(\left(\frac{1}{S} \right)^{K-1} \left(S-1 \right)^{|\mathcal{K}_i|-1}L+\left(\frac{1}{S} \right)^{K-1} \left(S-1 \right)^{|\mathcal{K}_i|}L \right)\frac{y_n}{S} \nonumber\\
&=\left(\frac{1}{S} \right)^{K-1} \left(S-1 \right)^{|\mathcal{K}_i|-1}L \left(\mathbf{B}_{\mathcal{P}}^T \mathbf{y} \right)_m\label{Incidence01G}\\
& \leq \left(\frac{1}{S} \right)^{K-1} \left(S-1 \right)^{|\mathcal{K}_i|-1}L \label{LinkLP1G} \\
&=\alpha_{i}, \label{BetaAlpha01G}
\end{align}
where in (\ref{Incidence01G}), $(\mathbf{x})_m$ denote the $m$-th element of the vector $\mathbf{x}$, (\ref{LinkLP1G}) follows because $\mathbf{y}$ satisfies the condition (\ref{Constraint02}) in (LP1), and (\ref{BetaAlpha01G}) follows by the definition of $\alpha_i$, which can be found immediately after (\ref{LongEquation}).  
Since the number of $(x_{\mathcal{K}_i^{[k]}}^{[k]}, x_{\mathcal{K}_i^{[k]} \bigcup \{\theta\}}^{[k]})$ retrieved from $\mathcal{T}_m$ is less than $\alpha_{i}$, $\mathbf{MDS}_{S\alpha_i\times\alpha_i} \left[\mathcal{I}_{\mathcal{K}_i^{[k]}}^{[k]m} \bigcup \mathcal{I}_{\mathcal{K}_i^{[k]} \bigcup \{\theta\}}^{[k]m},: \right]$ is full row rank for all $m \in [1:M]$, $k \in [1:K] \setminus \{\theta\}$, and $\mathcal{K}_i^{[k]}$, $i \in [1:\Delta]$. 

Thus, for all $m \in [1:M]$, $k \in [1:K] \setminus \{\theta\}$, the matrix 
\tiny{
\begin{align}
\mathbf{G}_k^m \triangleq
\begin{bmatrix}
\begin{matrix}
\mathbf{MDS}_{S\alpha_1\times\alpha_1}\left[\mathcal{I}_{\mathcal{K}_1^{[k]}}^{[k]m} \bigcup \mathcal{I}_{\mathcal{K}_1^{[k]} \bigcup \{\theta\}}^{[k]m},: \right] & \mathbf{0} & \mathbf{0} & \mathbf{0} \\ \hdashline[2pt/2pt]
\mathbf{0} & \mathbf{MDS}_{S\alpha_2\times\alpha_2}\left[\mathcal{I}_{\mathcal{K}_2^{[k]}}^{[k]m} \bigcup \mathcal{I}_{\mathcal{K}_2^{[k]} \bigcup \{\theta\}}^{[k]m},: \right] & \mathbf{0} & \mathbf{0} \\ \hdashline[2pt/2pt]
\mathbf{0} & \cdots & \ddots & \mathbf{0}\\ \hdashline[2pt/2pt]
\mathbf{0} & \mathbf{0} & \mathbf{0} & \mathbf{MDS}_{S \alpha_{\Delta}\times\alpha_{\Delta}}\left[\mathcal{I}_{\mathcal{K}_\Delta^{[k]}}^{[k]m} \bigcup \mathcal{I}_{\mathcal{K}_\Delta^{[k]} \bigcup \{\theta\}}^{[k]m},: \right]
\end{matrix}
\end{bmatrix} \nonumber
\end{align}
}
\normalsize
is full row rank. 
For notational convenience, let $\mathcal{I}^{[\theta]m}=\bigcup_{j=1}^\delta \mathcal{I}_{\mathcal{L}_j^{[\theta]}}^{[\theta]m}$, which is the indices of $x^{[\theta]}$ received by databases in $\mathcal{T}_m$. Similarly, define  $\mathcal{I}^{[k]m}=\bigcup_{i=1}^\Delta \left(\mathcal{I}_{\mathcal{K}_i^{[k]}}^{[k]m} \bigcup \mathcal{I}_{\mathcal{K}_i^{[k]}\bigcup \{\theta \}}^{[k]m}\right)$, $k \in [1:K] \setminus \{\theta\}$, as the set of indices of $x^{[k]}$ received by databases in $\mathcal{T}_m$ for the undesired message $k$. 
 Note that $\frac{L}{S}=\sum_{i=1}^\Delta \alpha_i \geq \left|\mathcal{I}^{[\theta]m} \right|=\left|\mathcal{I}^{[k]m} \right| \triangleq \tau_m$, $k \in [1:K] \setminus \{\theta\}$.
Databases in the colluding set $\mathcal{T}_m$ sees $\Big(x^{[\theta]}_{\mathcal{I}^{[\theta]m}} ,x^{[k]}_{\mathcal{I}^{[k]m}}, k \in [1:K] \setminus \{\theta\} \Big)$ with distribution $\Big(\mathbf{U}_\theta\left[\mathcal{I}^{[\theta]m},:\right]W_1, \mathbf{G}_k^m \mathbf{U}_{k}\left[\left(1:\frac{L}{S}\right),:\right]W_k, k \in [1:K] \setminus \{\theta\} \Big)$. To use Lemma \ref{AchPriv}, rewrite $\mathbf{U}_\theta \left[\mathcal{I}^{[\theta]m},: \right]=\begin{bmatrix} \mathbf{I}_{\tau_m} & \mathbf{0}_{\tau_m \times (\frac{L}{S}-\tau_m)} \end{bmatrix} \begin{bmatrix}\mathbf{U}_\theta \left[\mathcal{I}^{[\theta]m},:\right]\\
\mathbf{U}_\theta \left[ \mathcal{I}_c^{[\theta]m},: \right] \end{bmatrix}$, where $\mathcal{I}_c^{[\theta]m}$ is chosen as $\frac{L}{S}-\tau_m$ number of indices in $[1:L]$ who are not in $\mathcal{I}^{[\theta]m}$. Applying Lemma \ref{AchPriv}, we have
\begin{align}
&\Big(\mathbf{U}_\theta\left[\mathcal{I}^{[\theta]m},:\right], \mathbf{G}_k^m \mathbf{U}_{k}\left[\left(1:\frac{L}{S}\right),:\right], k \in [1:K] \setminus \{\theta\} \Big) \nonumber\\
& \sim  \left(\mathbf{U}_\theta[(1:\tau_m),:], \mathbf{U}_k[(1:\tau_m),:],k \in [1:K] \setminus \{\theta\} \right), \nonumber
\end{align}
which proves that the retrieval scheme is private. Thus, we have proved Theorem \ref{TheoremAchAnyY}.

\section{Proof of Theorem \ref{TheoremConverseAnyX}} \label{proof01}

For any PIR scheme, its rate, as defined in (\ref{DefineR}), satisfies
\begin{align}
    R_\mathcal{P} &= \frac{L}{\sum_{n=1}^{N}{H(A_n^{[\theta]})}} \nonumber\\
    &= \frac{L}{\sum_{n=1}^{N}{H(A_n^{[1]})}} \label{PrivacyAgain} \\
      & \leq \frac{L}{\sum_{n=1}^{N}{H(A_n^{[1]}|Q_{1:N}^{[1]})}}    \label{3-0-1} \\
      & \leq \frac{L}{H(A_{1:N}^{[1]}|Q_{1:N}^{[1]})},   \label{3-0-2}
\end{align}
where (\ref{PrivacyAgain}) is based on (\ref{PrivacyConstraint}), and   (\ref{3-0-1}) and (\ref{3-0-2}) are both due to conditioning reduces entropy. The following proof focuses on the lower bound of the denominator $H(A_{1:N}^{[1]}|Q_{1:N}^{[1]})$ in (\ref{3-0-2}). We have
\begin{align}
    H(A_{1:N}^{[1]}|Q_{1:N}^{[1]}) 
    & = H(A_{1:N}^{[1]}, W_1|Q_{1:N}^{[1]}) - H(W_1|A_{1:N}^{[1]}, Q_{1:N}^{[1]}) \nonumber\\
    & = H(A_{1:N}^{[1]}, W_1|Q_{1:N}^{[1]}) \label{conv01} \\
    & = H(W_1|Q_{1:N}^{[1]}) +H(A_{1:N}^{[1]}|W_1,Q_{1:N}^{[1]}) \nonumber\\
    &=L+H(A_{1:N}^{[1]}|W_1,Q_{1:N}^{[1]}), \label{conv06}
    \end{align}
where (\ref{conv01}) follows from (\ref{NoError}), and (\ref{conv06}) follows from (\ref{SysInd}) and (\ref{LL}). 
   
Now, we prove the following induction.
\begin{Lem} \label{LemmaInduction}
We have the following induction argument:
\begin{align}
S_2 H(A_{1:N}^{[k-1]}|W_{1:k-1}, Q_{1:N}^{[k-1]}) \geq L+H(A_{1:N}^{[k]}|W_{1:k}, Q_{1:N}^{[k]}), \quad k=2,3,\cdots,K. \nonumber
\end{align}
\end{Lem}
\begin{IEEEproof}
For each $\mathcal{T}_m \in \mathcal{P}$, $m=1,2,\cdots,M$, we may write
\begin{align}
  H(A_{1:N}^{[k-1]}|W_{1:k-1}, Q_{1:N}^{[k-1]})&= I(W_{k:K}; A_{1:N}^{[k-1]},Q_{1:N}^{[k-1]} |W_{1:k-1}) \label{MI01}\\
  & \geq  I(W_{k:K}; A_{\mathcal{T}_m}^{[k-1]},Q_{\mathcal{T}_m}^{[k-1]} |W_{1:k-1}) \nonumber\\
  &= H(A_{\mathcal{T}_m}^{[k-1]}|W_{1:k-1}, Q_{\mathcal{T}_m}^{[k-1]}) \label{MI02}\\
  &= H(A_{\mathcal{T}_m}^{[k]}|W_{1:k-1}, Q_{\mathcal{T}_m}^{[k]}) \label{3-0-3}\\
    &  \geq H(A_{\mathcal{T}_{m}}^{[k]}|W_{1:k-1}, Q_{1:N}^{[k]}), \label{conv07}
\end{align}
where (\ref{MI01}) and (\ref{MI02}) both follow from (\ref{SysInd}) and (\ref{DatabaseHonest}), and  (\ref{3-0-3}) follows from the privacy constraint in (\ref{PrivacyConstraint}).

We multiply the inequality derived from (\ref{MI01})-(\ref{conv07}) for each $\mathcal{T}_m$ on both sides by $x_m$, which is the $m$-th element of $\mathbf{x}$, and obtain
\begin{align}
x_m H(A_{1:N}^{[k-1]}|W_{1:k-1}, Q_{1:N}^{[k-1]}) \geq x_m H(A_{\mathcal{T}_{m}}^{[k]}|W_{1:k-1}, Q_{1:N}^{[k]}), \quad m=1,2,\cdots,M.  \label{Ping01}
\end{align}
Note that $\mathbf{x}$ satisfies (\ref{Positive02}) so we do not need to change the direction of the inequality in (\ref{Ping01}). 
Now, we add the $M$ inequalities denoted by (\ref{Ping01}) together and obtain (\ref{Ping02}), where we have used the definition of $S_2$, i.e., $S_2=\sum_{m=1}^M x_m$. 
The fact that $\mathbf{x}$ is rational and non-negative means that there exist non-negative integers $G_{\mathbf{x}}^1$, $G_{\mathbf{x}}^2$,$\cdots$, $G_{\mathbf{x}}^{M}$, $G_{\mathbf{x}}$, such that each $x_m$ can be expressed as 
\begin{align}
x_m=\frac{G_{\mathbf{x}}^m}{G_{\mathbf{x}}}, \quad \forall m \in [1: M]. \label{RationalInteger}
\end{align}
Thus, we have (\ref{Sub0101}). 

We write a more general summation than that on the right-hand side of (\ref{Sub0101}) as 
\begin{align}
\sum_{v=1}^V H(A_{\tilde{\mathcal{T}}_v}^{[k]}|W_{1:k-1}, Q_{1:N}^{[k]}), \label{GeneralSumApp}
\end{align} 
where $V$ is a positive integer, and $\tilde{\mathcal{T}}_v \subseteq [1:N]$, for $v \in [1:V]$. Note that the summation on the right-hand side of (\ref{Sub0101}) is a special case of (\ref{GeneralSumApp}) with $\tilde{\mathcal{T}}_v=\mathcal{T}_m$, where $m$ satisfies $\sum_{i=1}^{m-1} G_\mathbf{x}^i+1 \leq v \leq \sum_{i=1}^m G_\mathbf{x}^i$, and $V=\sum_{m=1}^M G_{\mathbf{x}}^m$, $m \in [1:M]$. We have the following definitions and a lemma regarding the sum (\ref{GeneralSumApp}). 
\begin{Def}
Define the subscript collection of the sum (\ref{GeneralSumApp}) as 
\begin{align}
\mathcal{A} \triangleq \{\tilde{\mathcal{T}}_1, \tilde{\mathcal{T}}_2, \cdots \tilde{\mathcal{T}}_V \}, \label{CorSets01}
\end{align}
where we have collected the subscript of $A_{\tilde{\mathcal{T}}_v}$ of sum (\ref{GeneralSumApp}) to form $\mathcal{A}$.
\end{Def}
\begin{Def}
We say that the sum (\ref{GeneralSumApp})  satisfies the \emph{even property} with the number $G$, if  
the number of times $n$ appears in its subscript collection $\mathcal{A}$ is equal to $G$ for each $n \in [1:N]$. 
\end{Def}
\begin{Lem} \label{NewNew}
When the sum (\ref{GeneralSumApp}) satisfies the even property with $G$, we have
\begin{align}
\sum_{v=1}^V H(A_{\tilde{\mathcal{T}}_v}^{[k]}|W_{1:k-1}, Q_{1:N}^{[k]}) \geq G \cdot H(A_{1:N}^{[k]}|W_{1:k-1},Q_{1:N}^{[k]}), \quad k=2,3,\cdots,K.  \label{Lemma4}
\end{align}
\end{Lem}
\begin{IEEEproof}
For a more fluent reading of the paper, we provide the details of the proof of Lemma \ref{NewNew} in Appendix \ref{ProofNewNew}, along with an illustrative example. The main idea is an iterated application of the sub-modular property of the entropy fuction\cite{schrijver2003combinatorial}. 
\end{IEEEproof}

Going back to the problem at hand, the subscript collection of the right-hand side of (\ref{Sub0101}) is
\begin{align}
\mathcal{A}_\mathbf{x} \triangleq \{\underbrace{\mathcal{T}_1, \cdots\mathcal{T}_1,}_{G_\mathbf{x}^1} \underbrace{\mathcal{T}_2, \cdots\mathcal{T}_2,}_{G_\mathbf{x}^2}\cdots, \underbrace{\mathcal{T}_M, \cdots \mathcal{T}_M}_{G_\mathbf{x}^M} \}.   \label{Ax}
\end{align}
The number of times $n$ appears in (\ref{Ax}) is $\sum_{j \in \mathcal{D}_n} G_\mathbf{x}^j$, where $\mathcal{D}_n$ is the set of indices of colluding sets in $\mathcal{P}$ which include Database $n$.

In the case of $\mathbf{B}_\mathcal{P} \mathbf{x}=\mathbf{1}_N$, the sum on the right-hand side of (\ref{Sub0101}) satisfies the even property with the number $G=G_\mathbf{x}$. This is because 
\begin{align}
1=\left(\mathbf{B}_\mathcal{P} \mathbf{x} \right)_n=\sum_{j \in \mathcal{D}_n}x_j=\sum_{j \in \mathcal{D}_n} \frac{G_\mathbf{x}^j}{G_\mathbf{x}}, \quad \forall n \in [1:N], \nonumber
\end{align}
where the last step follows from (\ref{RationalInteger}). 
Hence, applying Lemma \ref{NewNew}, we have proved (\ref{MoPing02})
in the case of $\mathbf{B}_\mathcal{P} \mathbf{x}=\mathbf{1}_N$. 

 In the case of $\mathbf{B}_\mathcal{P} \mathbf{x}>\mathbf{1}_N$, 
we have 
\begin{align}
1 \leq \left(\mathbf{B}_\mathcal{P} \mathbf{x} \right)_n=\sum_{j \in \mathcal{D}_n}x_j=\sum_{j \in \mathcal{D}_n} \frac{G_\mathbf{x}^j}{G_\mathbf{x}}, \quad \forall n \in [1:N]. \nonumber
\end{align}
We arbitrarily delete $\sum_{j \in \mathcal{D}_n} G_\mathbf{x}^j -G_\mathbf{x}$ number of $n$s from sets in $\mathcal{A}_\mathbf{x}$ of (\ref{Ax}), and obtain a new $\mathcal{A}$ as
 \begin{align}
 \{\underbrace{\mathcal{T}_1^{1}, \cdots\mathcal{T}_1^{G_\mathbf{x}^1},}_{G_\mathbf{x}^1} \underbrace{\mathcal{T}_2^{1}, \cdots\mathcal{T}_2^{G_\mathbf{x}^2},}_{G_\mathbf{x}^2}\cdots, \underbrace{\mathcal{T}_M^{1}, \cdots \mathcal{T}_M^{G_\mathbf{x}^M}}_{G_\mathbf{x}^M} \}. \label{ANewDefine}
 \end{align}
 Since we are deleting indices, we have $\mathcal{T}_m^{g} \subseteq \mathcal{T}_m$, $g \in [1:G_\mathbf{x}^m], m \in [1:M]$. Hence, the summation corresponding to (\ref{ANewDefine}) lower bounds the summation corresponding to (\ref{Ax}), i.e.,
 \begin{align}
&\sum_{m=1}^M G_\mathbf{x}^m  H(A_{\mathcal{T}_{m}}^{[k]}|W_{1:k-1}, Q_{1:N}^{[k]})
  \geq \sum_{m=1}^M \sum_{g=1}^{G_\mathbf{x}^m} H(A_{\mathcal{T}_{m}^{g}}^{[k]}|W_{1:k-1}, Q_{1:N}^{[k]}). \label{Property1Again}
 \end{align}
Since we have deleted $\sum_{j \in \mathcal{D}_n} G_\mathbf{x}^j -G_\mathbf{x}$ number of $n$s in $\mathcal{A}_\mathbf{x}$ of (\ref{Ax}), and obtained a new $\mathcal{A}$ as (\ref{ANewDefine}), the right-hand side of (\ref{Property1Again}) satisfies the even property with the number $G=G_\mathbf{x}$. Applying Lemma \ref{NewNew}, we have
\begin{align}
 \sum_{m=1}^M \sum_{g=1}^{G_\mathbf{x}^m} H(A_{\mathcal{T}_{m}^{g}}^{[k]}|W_{1:k-1}, Q_{1:N}^{[k]}) \geq G_\mathbf{x} \cdot H(A_{1:N}^{[k]}|W_{1:k-1},Q_{1:N}^{[k]}), \quad k=2,3,\cdots,K. \label{Case2}
\end{align}
From (\ref{Property1Again}) and (\ref{Case2}), we have (\ref{MoPing02}) for the case of $\mathbf{B}_\mathcal{P} \mathbf{x}>\mathbf{1}_N$ too. To make things more clear, we have included an example of the case $\mathbf{B}_\mathcal{P} \mathbf{x}>\mathbf{1}_N$ at the end of this subsection. 

Thus, when $\mathbf{x}$ satisfies $\mathbf{B}_\mathcal{P} \mathbf{x} \geq \mathbf{1}_N$, from (\ref{Sub0101}) and (\ref{MoPing02}), we have
\begin{align}
& \sum_{m=1}^M x_m  H(A_{\mathcal{T}_{m}}^{[k]}|W_{1:k-1}, Q_{1:N}^{[k]}) 
\geq  H(A_{1:N}^{[k]}|W_{1:k-1},Q_{1:N}^{[k]}), \quad k=2,3,\cdots,K.\label{submodular}
\end{align}

Finally, following from (\ref{Ping02}) and (\ref{submodular}), we have
\begin{align}
S_2 \cdot H(A_{1:N}^{[k-1]}|W_{1:k-1}, Q_{1:N}^{[k-1]}) &\geq H(A_{1:N}^{[k]}|W_{1:k-1},Q_{1:N}^{[k]}) \nonumber\\
&=H(A_{1:N}^{[k]}, W_k|W_{1:k-1},Q_{1:N}^{[k]}) \label{conv05}\\
&=H(W_k|W_{1:k-1}, Q_{1:N}^{[k]})+H(A_{1:N}^{[k]}|W_{1:k},Q_{1:N}^{[k]}) \nonumber\\
&=L+H(A_{1:N}^{[k]}|W_{1:k},Q_{1:N}^{[k]}), \nonumber
\end{align}
where (\ref{conv05}) follows from (\ref{NoError}). This proves Lemma \ref{LemmaInduction}. 
\end{IEEEproof}

Using (\ref{3-0-2}), (\ref{conv06}), Lemma \ref{LemmaInduction} and the fact that $H(A_{1:N}^{[K]}|W_{1:K}, Q_{1:N}^{[K]})=0$, we obtain the result of Theorem \ref{TheoremConverseAnyX}. 

\subsection{An example for the case of $\mathbf{B}_\mathcal{P} \mathbf{x}>\mathbf{1}_N$}
We provide the following example to illustrate how we delete indices to  obtain (\ref{ANewDefine}) and (\ref{Property1Again}) in the case of $\mathbf{B}_\mathcal{P} \mathbf{x}>\mathbf{1}_N$. 

For the collusion pattern of $N=5$ databases and $\mathcal{P}_5=\{\{1,3\}, \{2,3\}, \{3,4\}, \{1,5\}, \{2,5\}, \break \{4,5\}\}$, the corresponding incidence matrix is
\begin{align}
\mathcal{B}_{\mathcal{P}_5}=\left[ \begin{array}{c c c c c c}
1& 0& 0 &1&0&0 \\ 
0&1&0&0&1&0 \\
1&1&1&0&0&0\\
0&0&1&0&0&1  \\
0&0&0&1&1&1 \end{array} \right]. \nonumber
\end{align}
Solving linear programming problem (LP2), the optimal solution is
\begin{align}
\mathbf{x}^*=\left[ \begin{array}{c c c c c c c c c c c c} \frac{1}{2} & \frac{1}{2} &\frac{1}{2} &\frac{1}{2}&\frac{1}{2} &\frac{1}{2} \end{array} \right]^T, \label{OptimalX01}
\end{align}
and satisfies $\mathbf{B}_{\mathcal{P}_5} \mathbf{x}^*=\begin{bmatrix}1 & 1& \frac{3}{2} & 1 & \frac{3}{2}\end{bmatrix}^T >\mathbf{1}_N$. In the derivations below, we only use the fact that $\mathbf{x}^*$ is rational and feasible. We do not make use of the fact that $\mathbf{x}^*$ is optimal. 

For this example, with $\mathbf{x}^*$ in (\ref{OptimalX01}), we may pick $G_{\mathbf{x}^*}=2$, $G_{\mathbf{x}^*}^1=\cdots=G_{\mathbf{x}^*}^6=1$. The sum on the right-hand side of (\ref{Sub0101}) for this example becomes
\begin{align}
 &H(A_{\{1,3\}}^{[k]}|W_{1:k-1}, Q_{1:N}^{[k]})+H(A_{\{2,3\}}^{[k]}|W_{1:k-1}, Q_{1:N}^{[k]})+H(A_{\{3,4\}}^{[k]}|W_{1:k-1}, Q_{1:N}^{[k]})\nonumber\\
&+H(A_{\{1,5\}}^{[k]}|W_{1:k-1}, Q_{1:N}^{[k]})+H(A_{\{2,5\}}^{[k]}|W_{1:k-1}, Q_{1:N}^{[k]})+H(A_{\{4,5\}}^{[k]}|W_{1:k-1}, Q_{1:N}^{[k]}),  \label{ExampleSum}
\end{align}
and its associated subscript collection is 
\begin{align}
\{\{1,3\}, \{2,3\}, \{3,4\}, \{1,5\}, \{2,5\}, \{4,5\} \}. \label{ExampleSubscript}
\end{align}
Hence, the sum in (\ref{ExampleSum}) does not satisfy the even property, as Databases $1$,$2$ and $4$ appear $2$ times each, and Databases $3$ and $5$ appear $3$ times each. 

Arbitrarily delete $\sum_{j \in \mathcal{D}_3} G_j -G_\mathbf{x}=1$ number of 3, and  $\sum_{j \in \mathcal{D}_5} G_j -G_\mathbf{x}=1$ number of 5 from (\ref{ExampleSubscript}). There are $3 \times 3$ ways to do this, such as
\begin{align}
\{\{1\}, \{2,3\}, \{3,4\}, \{1\}, \{2,5\}, \{4,5\} \},  \label{ExampleDelete}
\end{align}
or $
\{\{1,3\}, \{2\}, \{3,4\}, \{1,5\}, \{2,5\}, \{4\} \}$ or $\{\{1\}, \{2,3\}, \{3,4\}, \{1,5\}, \{2,5\}, \{4\} \}$ etc. All 9 ways work for the following derivations and we take (\ref{ExampleDelete}) as an example. 

The subscript collection in (\ref{ExampleDelete})
corresponds to the sum
\begin{align}
 &H(A_{\{1\}}^{[k]}|W_{1:k-1}, Q_{1:N}^{[k]})+H(A_{\{2,3\}}^{[k]}|W_{1:k-1}, Q_{1:N}^{[k]})+H(A_{\{3,4\}}^{[k]}|W_{1:k-1}, Q_{1:N}^{[k]})\nonumber\\
&+H(A_{\{1\}}^{[k]}|W_{1:k-1}, Q_{1:N}^{[k]})+H(A_{\{2,5\}}^{[k]}|W_{1:k-1}, Q_{1:N}^{[k]})+H(A_{\{4,5\}}^{[k]}|W_{1:k-1}, Q_{1:N}^{[k]}),  \label{Weird01}
\end{align}
where it is easy to see that (\ref{Weird01}) is a lower bound to (\ref{ExampleSum}). Furthermore, (\ref{Weird01}) satisfies the even property, i.e., the number of times $n$ appears is 2 for $n \in [1:5]$. Hence, we have obtained a lower bound to (\ref{ExampleSum}), and this lower bound, i.e., (\ref{Weird01}), is a sum that satisfies the even property, and Lemma \ref{NewNew} may then be applied to the sum of (\ref{Weird01}).

\section{Proof of Lemma \ref{NewNew}} \label{ProofNewNew}
Consider the sum (\ref{GeneralSumApp}) and its corresponding subscript collection (\ref{CorSets01}), where 
the sum (\ref{GeneralSumApp}) satisfies the even property with the number $G$, i.e., 
the number of times $n$ appears in $\mathcal{A}$ of (\ref{CorSets01}) is equal to a number $G$ for each $n \in [1:N]$.

 The sub-modular property of the entropy function\cite{schrijver2003combinatorial} is
 \begin{align}
&H(A_{\mathcal{I}}^{[k]}|W_{1:k-1},Q_{1:N}^{[k]})+H(A_{\mathcal{J}}^{[k]}|W_{1:k-1},Q_{1:N}^{[k]}) \nonumber\\
& \geq H(A_{\mathcal{I} \bigcup \mathcal{J}}^{[k]}|W_{1:k-1},Q_{1:N}^{[k]})+H(A_{\mathcal{I} \bigcap \mathcal{J}}^{[k]}|W_{1:k-1},Q_{1:N}^{[k]}), \quad \mathcal{I}, \mathcal{J} \subseteq [1:N]. 
\label{Sub02}
\end{align}
After applying (\ref{Sub02}) once to the sum of two of the entropy terms of (\ref{GeneralSumApp}), the set $\mathcal{I}$ and $\mathcal{J}$ will be replaced by $\mathcal{I} \bigcup \mathcal{J}$ and $\mathcal{I} \bigcap \mathcal{J}$ in the lower bound of the sum. Correspondingly, sets $\mathcal{I}$ and $\mathcal{J}$ will be replaced by $\mathcal{I} \bigcup \mathcal{J}$ and $\mathcal{I} \bigcap \mathcal{J}$ in the subscript collection $\mathcal{A}$ associated with the new sum. 
%
Note that the number of times $n$, $n \in [1:N]$, appears in $(\mathcal{I},\mathcal{J})$ and in  $(\mathcal{I} \bigcup \mathcal{J},\mathcal{I} \bigcap \mathcal{J})$ is the same. Hence, 
the even property is always preserved after applying the sub-modular lower bounding of (\ref{Sub02}). Note also that if $\mathcal{I} \subseteq \mathcal{J}$ or $\mathcal{J} \subseteq \mathcal{I}$, then the lower bounding (\ref{Sub02}) becomes trivial. In particular, there is no need to apply (\ref{Sub02}) when $\mathcal{I}$ or $\mathcal{J}$ is the empty set or the whole set $[1:N]$. 

In the following, we propose an algorithm that iteratively applies the sub-modular lower bounding of (\ref{Sub02}) until the desired result, i.e., the right-hand side of (\ref{Lemma4}), is reached, see Algorithm \ref{AlgorithmSubmodular}. The feasibility and convergence of Algorithm \ref{AlgorithmSubmodular} will prove Lemma \ref{NewNew}. 

Algorithm \ref{AlgorithmSubmodular} is concerned with updating the subscript collection $\mathcal{A}$ after each lower bounding. Since there is a one-to-one correspondence between the subscript collection and its associated sum, Algorithm \ref{AlgorithmSubmodular} is in effect lower bounding the sum in (\ref{GeneralSumApp}) step by step to reach the right-hand side of (\ref{Lemma4}), which corresponds to the subscript collection $\mathcal{A}= \{\underbrace{[1:N], \cdots,[1:N],}_{G} \underbrace{\phi, \cdots \phi}_{V-G} \}$.  In each iteration, Algorithm \ref{AlgorithmSubmodular} first picks a set $a_1 \in \mathcal{A}$, that is neither the empty set or the whole set $[1:N]$, and is maximal, in the sense that no other set, except $[1:N]$ or a set equal to itself, contains it. Then, based on the $a_1$ picked, it picks another set $a_2 \in \mathcal{A}$ that is not a subset of $a_1$ nor the whole set $[1:N]$. 
Note that since $a_1$ is maximal, which will be proved in the convergence proof of Algorithm \ref{AlgorithmSubmodular}, we have that $a_1 \nsubseteq a_2$, and $a_2 \nsubseteq a_1$. 
Perform the sub-modular lower bounding of (\ref{Sub02}) for $\mathcal{I}=a_1, \mathcal{J}=a_2$ and update $\mathcal{A}$ corresponding to the new sum, where $a_1$ is replaced with the bigger set of $a_1 \bigcup a_2$ and $a_2$ is replaced with the smaller set of $a_1 \bigcap a_2$.  In the next iteration, if the new $a_1$ is not the whole set $[1:N]$ yet, use it again as $a_1$ and find a set $a_2 \in \mathcal{A}$ that is not a subset of $a_1$ nor the whole set $[1:N]$. Each iteration will make $a_1$ bigger and bigger until it becomes the whole set $[1:N]$, at which point, we pick another $a_1 \in \mathcal{A}$, that is neither the empty set or the whole set $[1:N]$, and is maximal, and start the iterations again.  The algorithm iterates until all sets left in $\mathcal{A}$ are either empty or the whole set, i.e., $[1:N]$. Since the sum (\ref{GeneralSumApp}) satisfies the even property with the number $G$, and with each iteration, the even property is continually satisfied, when the algorithm ends, the output has to be of the form $\mathcal{A}= \{\underbrace{[1:N], \cdots,[1:N],}_{G} \underbrace{\phi, \cdots \phi}_{V-G} \}$. 

 \hspace{-2in}
\begin{algorithm}
\label{AlgorithmSubmodular}
\DontPrintSemicolon
  
  \KwInput{$\mathcal{A} \triangleq \{\tilde{\mathcal{T}}_1, \tilde{\mathcal{T}}_2, \cdots \tilde{\mathcal{T}}_V \}$}
  \KwOutput{$\mathcal{A}= \{\underbrace{[1:N], \cdots,[1:N],}_{G} \underbrace{\phi, \cdots \phi}_{V-G} \}$}
   \Initialization{$a_1=\phi$}
   \While{$\exists a \in \mathcal{A}$ such that $a \neq [1:N] \text{ or } \phi$}
   {
   	Form a collection of sets $\mathcal{B}$ by removing the sets of $[1:N]$ and $\phi$ in $\mathcal{A}$\\
	If $a_1$ is not in $\mathcal{B}$, pick an $a_1 \in \mathcal{B}$ that is maximal, i.e., it satisfies: there does not exist a set $a' \in \mathcal{B}$ such that $a_1 \subsetneq a'$; Else, do nothing\\
	pick a set $a_2 \in \mathcal{B}$ where $a_2 \nsubseteq a_1$\\
	$b_1=a_1 \bigcup a_2$, $b_2=a_1 \bigcap a_2$\\
	In $\mathcal{A}$, replace $a_i$ with $b_i$, $i=1,2$\\
	$a_1=b_1, a_2=b_2$
	}
\caption{Lower bounding the sum in (\ref{GeneralSumApp}) to reach the right-hand side of (\ref{Lemma4})}
\end{algorithm}

\subsubsection{An Example illustrating Algorithm \ref{AlgorithmSubmodular}}: 

For the collusion pattern of $N=5$ databases and $\mathcal{P}_6=\{\{1,2\}, \{2,3,4\}, \{2,5\}, \{1,3,5\}, \{1,4,5\}, \break \{3,4,5\}\}$, the corresponding incidence matrix is
\begin{align}
\mathcal{B}_{\mathcal{P}_6}=\left[ \begin{array}{c c c c c c}
1& 0& 0 &1&1&0 \\ 
1&1&1&0&0&0 \\
0&1&0&1&0&1\\
0&1&0&0&1&1  \\
0&0&1&1&1&1 \end{array} \right]. \nonumber
\end{align}
Solving the linear programming problem (LP2), the optimal solution is
\begin{align}
\mathbf{x}^*=\left[ \begin{array}{c c c c c c c c c c c c} \frac{1}{5} & \frac{3}{5} &\frac{1}{5} &\frac{2}{5} &\frac{2}{5} &0 \end{array} \right]^T,  \label{OptimalX}
\end{align}
and satisfies $\mathbf{B}_{\mathcal{P}_6} \mathbf{x}^*=\mathbf{1}_N$. 
Hence, for this example, with the optimal $\mathbf{x}^*$ in (\ref{OptimalX}), we may pick $G_{\mathbf{x}^*}=5$, $G_{\mathbf{x}^*}^1=G_{\mathbf{x}^*}^3=1, G_{\mathbf{x}^*}^2=3, G_{\mathbf{x}^*}^4=G_{\mathbf{x}^*}^5=2, G_{\mathbf{x}^*}^6=0$. In the following, we do not use the optimality of $\mathbf{x}^*$, only that it generates a sum that satisfies the even property. 

The right-hand side of (\ref{Sub0101}) becomes
\begin{align}
&H(A_{\{1,2\}}^{[k]}|W_{1:k-1}, Q_{1:N}^{[k]})+3H(A_{\{2,3,4\}}^{[k]}|W_{1:k-1}, Q_{1:N}^{[k]})+H(A_{\{2,5\}}^{[k]}|W_{1:k-1}, Q_{1:N}^{[k]})\nonumber\\
&+2H(A_{\{1,3,5\}}^{[k]}|W_{1:k-1}, Q_{1:N}^{[k]})+2H(A_{\{1,4,5\}}^{[k]}|W_{1:k-1}, Q_{1:N}^{[k]}).  \label{NoGeneral}
\end{align}
Written in the form of (\ref{GeneralSumApp}), we have
\begin{align}
(\ref{NoGeneral}) = &\underline{H(A_{\{1,2\}}^{[k]}|W_{1:k-1}, Q_{1:N}^{[k]})}+\underline{H(A_{\{2,3,4\}}^{[k]}|W_{1:k-1}, Q_{1:N}^{[k]})}+H(A_{\{2,3,4\}}^{[k]}|W_{1:k-1}, Q_{1:N}^{[k]})\nonumber\\
&+H(A_{\{2,3,4\}}^{[k]}|W_{1:k-1}, Q_{1:N}^{[k]})+H(A_{\{2,5\}}^{[k]}|W_{1:k-1}, Q_{1:N}^{[k]})+H(A_{\{1,3,5\}}^{[k]}|W_{1:k-1}, Q_{1:N}^{[k]}) \nonumber\\
&+H(A_{\{1,3,5\}}^{[k]}|W_{1:k-1}, Q_{1:N}^{[k]})+H(A_{\{1,4,5\}}^{[k]}|W_{1:k-1}, Q_{1:N}^{[k]})+H(A_{\{1,4,5\}}^{[k]}|W_{1:k-1}, Q_{1:N}^{[k]}), \label{Iteration01}
\end{align}
and the corresponding input to Algorithm 1 is $\mathcal{A}=\{\{1,2\}, \{2,3,4\}, \{2,3,4\}, \{2,3,4\}, \{2,5\},  \break \{1,3,5\}, \{1,3,5\}, \{1,4,5\}, \{1,4,5\}\}$. Note that the number of times $n$ appears in $\mathcal{A}$ is $5$, for $n \in [1:5]$. Thus, the summation in (\ref{Iteration01}) satisfies the even property.

In the first iteration, $\mathcal{B}=\mathcal{A}$ in Step 2 of Algorithm 1. In Step 3, Pick $a_1=\{1,2\} \in \mathcal{B}$ which is maximal and further pick $a_2=\{2,3,4\} \in \mathcal{B}$ in Step 4 which is not a subset of $a_1$. 
Applying the sub-modular property of the entropy function on the two underlined terms in (\ref{Iteration01}), we have
\begin{align}
(\ref{Iteration01}) \geq &\underline{H(A_{\{1,2,3,4\}}^{[k]}|W_{1:k-1}, Q_{1:N}^{[k]})}+H(A_{\{2\}}^{[k]}|W_{1:k-1}, Q_{1:N}^{[k]})+H(A_{\{2,3,4\}}^{[k]}|W_{1:k-1}, Q_{1:N}^{[k]})\nonumber\\
&+H(A_{\{2,3,4\}}^{[k]}|W_{1:k-1}, Q_{1:N}^{[k]})+\underline{H(A_{\{2,5\}}^{[k]}|W_{1:k-1}, Q_{1:N}^{[k]})}+H(A_{\{1,3,5\}}^{[k]}|W_{1:k-1}, Q_{1:N}^{[k]}) \nonumber\\
&+H(A_{\{1,3,5\}}^{[k]}|W_{1:k-1}, Q_{1:N}^{[k]})+H(A_{\{1,4,5\}}^{[k]}|W_{1:k-1}, Q_{1:N}^{[k]})+H(A_{\{1,4,5\}}^{[k]}|W_{1:k-1}, Q_{1:N}^{[k]}), \label{Iteration02}
\end{align}
which corresponds to the new $\mathcal{A}=\{\{1,2,3,4\}, \{2\}, \{2,3,4\}, \{2,3,4\}, \{2,5\},   \{1,3,5\}, \{1,3,5\}, \break \{1,4,5\}, \{1,4,5\}\}$. The new $a_1=\{1,2,3,4\}$ and $a_2=\{2\}$. Note that the number of times $n$ appears in $\mathcal{A}$ is still $5$, for $n \in [1:5]$. Thus, the summation in (\ref{Iteration02}) also satisfies the even property.

In Iteration 2, $\mathcal{B}=\mathcal{A}$ in Step 2 of Algorithm 1. In Step 3, since $a_1=\{1,2,3,4\}$ is not the whole set and thus in $\mathcal{B}$, we continue using this as $a_1$, and further pick $a_2=\{2,5\}$ in Step 4 which is not a subset of $a_1$. 
Applying the sub-modular property of the entropy function on the two underlined terms in (\ref{Iteration02}), we have
\begin{align}
(\ref{Iteration02}) \geq &H(A_{[1:5]}^{[k]}|W_{1:k-1}, Q_{1:N}^{[k]})+H(A_{\{2\}}^{[k]}|W_{1:k-1}, Q_{1:N}^{[k]})+\underline{H(A_{\{2,3,4\}}^{[k]}|W_{1:k-1}, Q_{1:N}^{[k]})}\nonumber\\
&+H(A_{\{2,3,4\}}^{[k]}|W_{1:k-1}, Q_{1:N}^{[k]})+H(A_{\{2\}}^{[k]}|W_{1:k-1}, Q_{1:N}^{[k]})+\underline{H(A_{\{1,3,5\}}^{[k]}|W_{1:k-1}, Q_{1:N}^{[k]})} \nonumber\\
&+H(A_{\{1,3,5\}}^{[k]}|W_{1:k-1}, Q_{1:N}^{[k]})+H(A_{\{1,4,5\}}^{[k]}|W_{1:k-1}, Q_{1:N}^{[k]})+H(A_{\{1,4,5\}}^{[k]}|W_{1:k-1}, Q_{1:N}^{[k]}), \label{Iteration03}
\end{align}
which corresponds to the new $\mathcal{A}=\{[1:5], \{2\}, \{2,3,4\}, \{2,3,4\}, \{2\},   \{1,3,5\}, \{1,3,5\},  \break \{1,4,5\}, \{1,4,5\}\}$. The new $a_1=[1:5]$ and $a_2=\{2\}$. Again, the number of times $n$ appears in $\mathcal{A}$ is still $5$, for $n \in [1:5]$. Thus, the summation in (\ref{Iteration03}) also satisfies the even property. 

In Iteration 3, $\mathcal{B}=\{\{2\}, \{2,3,4\}, \{2,3,4\}, \{2\},   \{1,3,5\}, \{1,3,5\},  \{1,4,5\}, \{1,4,5\}\}$ in Step 2 of Algorithm 1. In Step 3, since $a_1=[1:5]$ is the whole set and thus not in $\mathcal{B}$, we pick another $a_1=\{2,3,4\}$ which is maximal. Further pick $a_2=\{1,3,5\}$ in Step 4 which is not a subset of $a_1$. 
Applying the sub-modular property of the entropy function on the two underlined terms in (\ref{Iteration03}), we have
\begin{align}
(\ref{Iteration03})\geq &H(A_{[1:5]}^{[k]}|W_{1:k-1}, Q_{1:N}^{[k]})+H(A_{\{2\}}^{[k]}|W_{1:k-1}, Q_{1:N}^{[k]})+H(A_{[1:5]}^{[k]}|W_{1:k-1}, Q_{1:N}^{[k]})\nonumber\\
&+\underline{H(A_{\{2,3,4\}}^{[k]}|W_{1:k-1}, Q_{1:N}^{[k]})}+H(A_{\{2\}}^{[k]}|W_{1:k-1}, Q_{1:N}^{[k]})+H(A_{\{3\}}^{[k]}|W_{1:k-1}, Q_{1:N}^{[k]}) \nonumber\\
&+\underline{H(A_{\{1,3,5\}}^{[k]}|W_{1:k-1}, Q_{1:N}^{[k]})}+H(A_{\{1,4,5\}}^{[k]}|W_{1:k-1}, Q_{1:N}^{[k]})+H(A_{\{1,4,5\}}^{[k]}|W_{1:k-1}, Q_{1:N}^{[k]}), \label{Iteration04}
\end{align}
which corresponds to the new $\mathcal{A}=\{[1:5], \{2\}, [1:5], \underline{\{2,3,4\}}, \{2\},   \{3\}, \overline{\{1,3,5\}},   \{1,4,5\}, \break \{1,4,5\}\}$. The new $a_1=[1:5]$ and $a_2=\{3\}$. Again, the number of times $n$ appears in $\mathcal{A}$ is still $5$, for $n \in [1:5]$. Thus, the summation in (\ref{Iteration04}) also satisfies the even property.

We carry on like this with the following iterations, and at the end each iteration, we have the updated subscript collection $\mathcal{A}$ being
\begin{align}
\text{Iteration 4:} \quad&\mathcal{A}=\{[1:5], \underline{\{2\}}, [1:5], [1:5], \{2\},  \overline{ \{3\}}, \{3\},   \{1,4,5\}, \{1,4,5\}\}; \nonumber\\
\text{Iteration 5:} \quad &\mathcal{A}=\{[1:5], \underline{\{2,3\}}, [1:5], [1:5], \{2\},  \phi, \{3\},  \overline{ \{1,4,5\}}, \{1,4,5\}\};\nonumber\\
\text{Iteration 6:} \quad &\mathcal{A}=\{[1:5], [1:5], [1:5], [1:5], \underline{\{2\}},  \phi, \overline{\{3\}},  \phi, \{1,4,5\}\}; \nonumber\\
\text{Iteration 7:}\quad &\mathcal{A}=\{[1:5], [1:5], [1:5], [1:5], \underline{\{2,3\}},  \phi, \phi,  \phi, \overline{\{1,4,5\}}\};  \nonumber\\
\text{Iteration 8:}\quad &\mathcal{A}=\{[1:5], [1:5], [1:5], [1:5], [1:5],  \phi, \phi,  \phi, \phi\}, \nonumber
\end{align}
where the underlined set is $a_1$ in the next iteration and the overlined set is what we pick to be $a_2$, which is not a subset of $a_1$, in the next iteration. This corresponds to the following derivations:
\begin{align}
(\ref{Iteration04}) \geq &H(A_{[1:5]}^{[k]}|W_{1:k-1}, Q_{1:N}^{[k]})+H(A_{\{2\}}^{[k]}|W_{1:k-1}, Q_{1:N}^{[k]})+H(A_{[1:5]}^{[k]}|W_{1:k-1}, Q_{1:N}^{[k]})\nonumber\\
&+H(A_{[1:5]}^{[k]}|W_{1:k-1}, Q_{1:N}^{[k]})+H(A_{\{2\}}^{[k]}|W_{1:k-1}, Q_{1:N}^{[k]})+H(A_{\{3\}}^{[k]}|W_{1:k-1}, Q_{1:N}^{[k]}) \nonumber\\
&+H(A_{\{3\}}^{[k]}|W_{1:k-1}, Q_{1:N}^{[k]})+H(A_{\{1,4,5\}}^{[k]}|W_{1:k-1}, Q_{1:N}^{[k]})+H(A_{\{1,4,5\}}^{[k]}|W_{1:k-1}, Q_{1:N}^{[k]}) \nonumber\\
\geq &H(A_{[1:5]}^{[k]}|W_{1:k-1}, Q_{1:N}^{[k]})+H(A_{\{2,3\}}^{[k]}|W_{1:k-1}, Q_{1:N}^{[k]})+H(A_{[1:5]}^{[k]}|W_{1:k-1}, Q_{1:N}^{[k]})\nonumber\\
&+H(A_{[1:5]}^{[k]}|W_{1:k-1}, Q_{1:N}^{[k]})+H(A_{\{2\}}^{[k]}|W_{1:k-1}, Q_{1:N}^{[k]})+0 \nonumber\\
&+H(A_{\{3\}}^{[k]}|W_{1:k-1}, Q_{1:N}^{[k]})+H(A_{\{1,4,5\}}^{[k]}|W_{1:k-1}, Q_{1:N}^{[k]})+H(A_{\{1,4,5\}}^{[k]}|W_{1:k-1}, Q_{1:N}^{[k]})  \nonumber\\
\geq &H(A_{[1:5]}^{[k]}|W_{1:k-1}, Q_{1:N}^{[k]})+H(A_{[1:5]}^{[k]}|W_{1:k-1}, Q_{1:N}^{[k]})+H(A_{[1:5]}^{[k]}|W_{1:k-1}, Q_{1:N}^{[k]})\nonumber\\
&+H(A_{[1:5]}^{[k]}|W_{1:k-1}, Q_{1:N}^{[k]})+H(A_{\{2\}}^{[k]}|W_{1:k-1}, Q_{1:N}^{[k]})+0 \nonumber\\
&+H(A_{\{3\}}^{[k]}|W_{1:k-1}, Q_{1:N}^{[k]})+0+H(A_{\{1,4,5\}}^{[k]}|W_{1:k-1}, Q_{1:N}^{[k]})  \nonumber\\
\geq & \cdots \nonumber\\
\geq & 5 H(A_{\{[1:5]\}}^{[k]}|W_{1:k-1}, Q_{1:N}^{[k]}). \nonumber
\end{align}
Note that the even property is preserved in each step of the lower bounding.

\subsubsection{Feasibility proof of Algorithm \ref{AlgorithmSubmodular}}
 Algorithm feasibility depends on whether we are able to find the required sets in Steps 3 and 4. In Step 3, if $a_1$ from the previous iteration is not the whole set $[1:N]$, we will continue to use it as $a_1$ in the next iteration. In the case where $a_1$ from the previous iteration \emph{is} the whole set $[1:N]$, we need to pick another $a_1 \in \mathcal{B}$ that is maximal. We prove that such an $a_1$ can always be found by contradiction: suppose no such $a_1 \in \mathcal{B}$ is found, which means that for every set in $\mathcal{B}$, you can find another set that strictly contains it. But this can not be true as the number of sets in $\mathcal{B}$ is finite. Thus, we have proved that we can always find a set $a_1$ that satisfies the condition in Step 3.

We will prove the feasibility of Step 4 by contradiction: suppose no such set $a_2$ can be found, which means that all the other sets in $\mathcal{B}$ are 
 either the same as $a_1$, or a subset of $a_1$. We know that $a_1$ is not the whole set because it belongs to $\mathcal{B}$, so there is at least one index $j_0 \in [1:N]$, such that $j_0 \notin a_1$. Also, $a_1$ is not the empty set, so there exists at least one index $j_1 \in [1: N]$, such that $j_1 \in a_1$. The number of times $j_0$ appears in $\mathcal{B}$ is zero as $\mathcal{B}$ only contains sets that are either the same as $a_1$, or a subset of $a_1$. But the number of times $j_1$ appears in $\mathcal{B}$ is at least 1, as it is contained in $a_1$. Hence, the number of times $j_0$ appears in $\mathcal{B}$ is strictly less than the number of times $j_1$ appears in $\mathcal{B}$, and as a consequence, the number of times $j_0$ appears in $\mathcal{A}$ is strictly less than the number of times $j_1$ appears in $\mathcal{A}$, violating the even property. However, the even property should always hold when applying the sub-modular lower bounding of (\ref{Sub02}), which means that it is satisfied at each iteration of the algorithm. Thus, we have arrived at a contradiction, which means that the $a_2$ in Step 4 can always be found. 

\subsubsection{Convergence proof of Algorithm \ref{AlgorithmSubmodular}}
First, we claim that Step 3 always gives us a set $a_1$, that is maximal. We start the iteration by picking an $a_1 \in \mathcal{B}$ that is maximal. As the iterations goes on, it becomes bigger and bigger while the other sets either remain the same or become smaller. Hence, it will remain maximal in each iteration until it reaches the whole set, $[1:N]$, at which point, a new maximal $a_1$ will be picked and it will also remain maximal during future iterations and so on and so forth. So Step 3 always gives us a set $a_1$ which is maximal. 

Since $a_1$ is maximal in each iteration, in Step 4, we can not find an $a$ that strictly contains $a_1$. Since Step 4 also ensures that we do not pick an $a_2$ that is a subset of $a_1$, we have $a_1 \nsubseteq a_2$ and $a_2 \nsubseteq a_1$ in each iteration. Thus, the size of $a_1$ gets increased by at least one in each iteration. So each maximal set picked from $\mathcal{B}$ takes at most $N-1$
iterations to reach the whole set $[1:N]$. Since there are at most $V$  sets in $\mathcal{A}$, Algorithm \ref{AlgorithmSubmodular} will stop after at most $(N-1)V$ iterations. 

By proving the feasibility and convergence of Algorithm \ref{AlgorithmSubmodular}, we have shown that the algorithm can indeed change  $\mathcal{A} \triangleq \{\tilde{\mathcal{T}}_1, \tilde{\mathcal{T}}_2, \cdots \tilde{\mathcal{T}}_V \}$ step by step to
$\mathcal{A}= \{\underbrace{[1:N], \cdots,[1:N],}_{G} \underbrace{\phi, \cdots \phi}_{V-G} \}$, which means that we can lower-bound the sum of (\ref{GeneralSumApp}) to reach the right-hand side of (\ref{Lemma4}). 

Thus, Lemma \ref{NewNew} is proved.

\bibliographystyle{unsrt}
\bibliography{BPIR}

\end{document}